\documentclass[a4paper,oneside,reqno]{amsart}
\usepackage[top=1.5in, bottom=1in, left=1in, right=1in]{geometry}
\usepackage{cite}
\usepackage{amsmath,amssymb,amsfonts,amsthm}
\usepackage{algorithmic}
\usepackage{graphicx}
\usepackage{algorithm,algorithmic}
\usepackage{hyperref}
\hypersetup{hidelinks=true}
\usepackage{textcomp}

\usepackage{balance}
\usepackage{subcaption}
\usepackage{placeins}
\usepackage[binary-units]{siunitx}

\makeatletter
\let\NAT@parse\undefined
\makeatother
\usepackage[sort&compress,numbers]{natbib}

\usepackage{xcolor}
\definecolor{cblue}{HTML}{045275}
\definecolor{cred}{HTML}{F0746E}
\definecolor{cgreen}{HTML}{7CCBA2}
\definecolor{legreddraw}{HTML}{E41A1C}
\definecolor{legbluedraw}{HTML}{377EB8}
\definecolor{leggreendraw}{HTML}{4DAF4A}
\definecolor{legvioletdraw}{HTML}{984EA3}
\definecolor{legredfill}{HTML}{B61516}
\definecolor{legbluefill}{HTML}{2C6593}
\definecolor{leggreenfill}{HTML}{3E8C3B}
\definecolor{legvioletfill}{HTML}{7A3E82}

\usepackage{amsmath, amssymb, amsfonts, mathtools, bm}
\usepackage{stmaryrd} 
\usepackage{mleftright}
\usepackage{mymath} 
\usepackage{arydshln} 

\usepackage{parskip} 
\usepackage{tabularx,ragged2e,booktabs,caption,multirow} 
\usepackage{array}
\usepackage{makecell} 

\usepackage{extarrows}
\RequirePackage{rotating} 
\RequirePackage{fix-cm}
\usepackage{stackengine}
\usepackage{multicol}
\usepackage{rotating}
\usepackage{booktabs}

\usepackage{pgfplots}
\usepgfplotslibrary{patchplots}
\usepgfplotslibrary{colormaps}
\pgfplotsset{compat=newest}
\pgfplotsset{
  layers/axis lines on top/.define layer set={
    axis background,
    axis grid,
    axis ticks,
    axis tick labels,
    pre main,
    main,
    axis lines,
    axis descriptions,
    axis foreground,
  }{/pgfplots/layers/standard},
}
\usepgfplotslibrary{fillbetween}
\usepgfplotslibrary{colorbrewer}

\usepackage{tikz}
\usetikzlibrary{intersections}
\usetikzlibrary{shapes, arrows, shapes.misc, arrows.meta, positioning, matrix, calc, fit, fadings, patterns}
\usetikzlibrary{calc,patterns,decorations.pathmorphing,decorations.markings,arrows, arrows.meta,pgfplots.colorbrewer}
\usetikzlibrary{shapes, arrows, arrows.meta, positioning, calc,pgfplots.colorbrewer}
\newlength{\Llong}
\newlength{\Lmid}
\newlength{\Lshort}
\usepackage{csvsimple}
\usetikzlibrary{external}
\newcommand{\inputtikz}[1]{%
\centering
\includegraphics[scale=1]{figures/#1.pdf}
}
\pgfdeclarelayer{bg}    
\pgfsetlayers{bg,main}  
\usetikzlibrary{plotmarks}
\newenvironment{customlegend}[1][]{%
    \begingroup
    \csname pgfplots@init@cleared@structures\endcsname
    \pgfplotsset{#1}%
}{%
    \csname pgfplots@createlegend\endcsname
    \endgroup
}%
\def\addlegendimage{\csname pgfplots@addlegendimage\endcsname}
\newcommand{\addlegendimageintext}[1]{%
    \tikzexternaldisable
    \tikz {
        \begin{customlegend}[
            legend entries={\empty},
            legend style={
                at={(current bounding box.north west)},
    			anchor=north west,
                draw=none,
                inner sep=1.5pt,
                column sep=0pt,
                nodes={inner sep=0pt}}]
        \addlegendimage{#1}
        \end{customlegend}
    }%
    \tikzexternalenable
}

\usepackage[normalem]{ulem}

\newtheorem{lemma}{Lemma}

\newtheorem{theorem}{Theorem}

\newtheorem{remark}{Remark}

\pgfplotsset{colormap={CMsurf}{color=(white) color=(legbluedraw) color=(legvioletdraw) color=(legreddraw) color=(yellow)}}

\pgfplotsset{colormap={CMsurfrev}{color=(yellow) color=(legreddraw) color=(legvioletdraw) color=(legbluedraw) color=(white)}}

\pgfplotsset{colormap={CMsurfr}{color=(yellow) color=(legreddraw) color=(legvioletdraw) color=(legbluedraw) color=(white)}}

\tikzset{
    png export/.style={
        external/system call/.add={}{; convert -density 150 "\image.pdf" -quality 90 "\image.png"},
        /pgf/images/external info,
        /pgf/images/include external/.code={
            \includegraphics[width=\pgfexternalwidth,height=\pgfexternalheight]{##1.png}
        },
    }
}

\usepackage{xcolor,calc}

\begin{document}
\title[Control of Cross-Directional Systems using the GSVD]{Control of Cross-Directional Systems using the\\Generalised Singular Value Decomposition}
\author{Idris Kempf}
\author{Paul J. Goulart}
\author{Stephen Duncan}
\thanks{This work was supported in part by the Diamond Light Source and in part by the Engineering and Physical Sciences Research Council (EPSRC) with a Collaborative Awards in Science and Engineering (CASE) studentship.}
\thanks{The authors are with the Department of Engineering Science, University of Oxford, Oxford OX1 3PJ, UK (e-mails: idris.kempf@eng.ox.ac.uk, paul.goulart@eng.ox.ac.uk, stephen.duncan@eng.ox.ac.uk).}


\maketitle

\begin{abstract}
Diamond Light Source produces synchrotron radiation by accelerating electrons to relativistic speeds. In order to maximise the intensity of the radiation, vibrations of the electron beam are attenuated by a multi-input multi-output (MIMO) control system actuating hundreds of magnets at kilohertz rates. For future accelerator configurations, in which two separate arrays of magnets with different bandwidths are used in combination, standard accelerator control design methods based on the singular value decomposition (SVD) of the system gain matrix are not suitable. We therefore propose to use the generalised singular value decomposition (GSVD) to decouple a two-array cross-directional (CD) system into sets of two-input single-output (TISO) and single-input single-output (SISO) systems. We demonstrate that the two-array decomposition is linked to a single-array system, which is used to accommodate ill-conditioned systems and compensate for the non-orthogonality of the GSVD. The GSVD-based design is implemented and validated through real-world experiments at Diamond. Our approach provides a natural extension of single-array methods and has potential application in other CD systems, including paper making, steel rolling or battery manufacturing processes.
\end{abstract}


\section{Introduction}\label{sec:introduction}
Diamond Light Source (Diamond) is the UK's national synchrotron facility that produces \emph{synchrotron radiation} for scientific and commercial research.  Synchrotron radiation is used for various experimental techniques that allow materials and organisms to be characterised at atomic scales. The radiation is generated by electrons moving at relativistic speeds around a \SI{500}{\meter} circumference \emph{storage ring} and a key feature of synchrotron light sources is the exceptional intensity of the synchrotron radiation. However, the radiation intensity is significantly reduced by vibrations of the electron beam in the horizontal and vertical planes perpendicular to the direction of motion. The coupling between the planes is negligible so the horizontal and vertical vibrations can be treated as two separate control problems~\cite{SANDIRAWINDUP}. To attenuate these vibrations, a \emph{fast orbit feedback} (FOFB) system is employed, using up to $172$ \emph{beam position monitors} (BPMs) as sensors and up to $173$ \emph{corrector magnets} as inputs. The FOFB system operates at a sampling rate of \SI{10}{\kHz} and reduces the root-mean-square deviation of the electrons to \SI{10}{\percent} of the beam size up to a closed-loop bandwidth of \SI{226}{\Hz}. The electron beam dynamics are modelled by a linear time-invariant \emph{cross-directional} (CD) system~\cite{SANDIRAWINDUP}:
\begin{equation}\label{eq:CDsystem}
y(s)=Rg(s)u(s)+ d(s),
\end{equation}
where $s\in\C$ is the Laplace variable, $u:\C\mapsto\C^{n_u}$ are the inputs, $y:\C\mapsto\C^{n_y}$ the outputs and $d:\C\mapsto\C^{n_y}$ the disturbances. The constant matrix $\inR{R}{n_y}{n_u}$ is referred to as the \emph{orbit response matrix}~\cite{CROSSDIR}. The stable transfer function $g:\C\mapsto\C$ captures the temporal dynamics of the actuators. In addition to the high sampling frequency and the large number of inputs and outputs, the control problem is aggravated by the large condition number of $R$, typically between $10^3$ and $10^4$ for large synchrotron facilities like Diamond.

Other examples of CD systems are found in industrial applications, including paper making, metal rolling and polymer film extrusion~\cite{CROSSDIR}. Analogous to synchrotron light sources, these applications involve large-scale CD systems for which optimisation-based synthesis, such as $\mathcal{H}_\infty$ or $\mathcal{H}_2$ control~\cite[Ch.\ 9]{SKOGESTADMULTI}, can be difficult to implement~\cite{CROSSDIR}. Moreover, for high sampling rates, controllers that involve real-time optimisation are difficult to realise in practice~\cite{SANDIRAWINDUP}. However, the controller synthesis and analysis of CD systems can be greatly simplified using a \emph{modal transformation} that decouples the multi-input multi-output (MIMO) system~\eqref{eq:CDsystem} by substituting the singular value decomposition (SVD) $R=U\Sigma\trans{V}$. In modal space, the MIMO system reduces to a set of independent single-input, single-output (SISO) systems, allowing mode-by-mode controller design.

To meet users' demand for enhanced experimental facilities, Diamond is upgrading their facility to a next-generation light source, Diamond-II. As a result, the target for the root mean square deviation has been tightened from \SI{10}{\percent} to \SI{3}{\percent} of beam size up to a bandwidth of \SI{1}{\kHz}. To meet these requirements, the number of sensors and inputs is increased and an additional array of actuators introduced, i.e.\
\begin{align}\label{eq:system}
y(s) = R_\mathrm{s} g_\mathrm{s}(s) u_\mathrm{s}(s) + R_\mathrm{f} g_\mathrm{f}(s) u_\mathrm{f}(s) + d(s),
\end{align}
where $\inR{R_\mathrm{s}}{n_y}{n_\mathrm{s}}$, $\inR{R_\mathrm{f}}{n_y}{n_\mathrm{f}}$, $n_y=n_\mathrm{s}=252$, $n_\mathrm{f}=144$, and the subscripts ``$\mathrm{s}$'' and ``$\mathrm{f}$'' refer to \emph{slow} and \emph{fast}. By introducing a second array of actuators, the control effort is split onto ``slow'' magnets with a low bandwidth and a strong magnetic field, and ``fast'' magnets with a high bandwidth and a weaker magnetic field. 

Analogous to the single-array case, decoupling~\eqref{eq:system} facilitates the controller synthesis, but standard modal decomposition \emph{cannot} be applied when two or more actuator arrays are present. To see why, substitute the standard SVDs $R_{(\cdot)}=\bar{U}_{(\cdot)}\bar{\Sigma}_{(\cdot)}\bar{V}_{(\cdot)}^\Tr$, where ${(\cdot)}=\lbrace\mathrm{s},\mathrm{f}\rbrace$, in~\eqref{eq:system} to obtain
\begin{align*}
y(s) = \bar{U}_\mathrm{s}\bar{\Sigma}_\mathrm{s}\bar{V}_\mathrm{s}^\Tr g_\mathrm{s}(s) u_\mathrm{s}(s) + \bar{U}_\mathrm{f}\bar{\Sigma}_\mathrm{f}\bar{V}_\mathrm{f}^\Tr g_\mathrm{f}(s) u_\mathrm{f}(s) + d(s).
\end{align*}
Left-multiplying with $\bar{U}_\mathrm{s}^\Tr$ and defining $\bar{y}(s)\eqdef\bar{U}_\mathrm{s}^\Tr y(s)$, $\bar{u}_{(\cdot)}(s)\eqdef\bar{V}_{(\cdot)}^\Tr u_{(\cdot)}(s)$, and $\bar{d}(s)\eqdef\bar{U}_\mathrm{s}^\Tr d(s)$ yields
\begin{align}
\bar{y}(s) = \bar{\Sigma}_\mathrm{s} g_\mathrm{s}(s) \bar{u}_\mathrm{s}(s) + \bar{U}_\mathrm{s}^\Tr\bar{U}_\mathrm{f}\bar{\Sigma}_\mathrm{f} g_\mathrm{f}(s) \bar{u}_\mathrm{f}(s) + \bar{d}(s),
\end{align}
which shows that, using the standard SVDs of $R_{(\cdot)}$, system~\eqref{eq:system} is decoupled iff $\bar{U}_\mathrm{s}^\Tr\bar{U}_\mathrm{f}=I$.

Although extensions of modal decomposition have been proposed~\cite{STEPHENMULTIARR2,SANDIRAMULTIARRAY}, they rely on the analysis of the controllable subspaces of the slow and fast actuators arrays, which for system~\eqref{eq:system} are $\mathcal{Y}_\mathrm{s}\eqdef\range(R_\mathrm{s})\subseteq\R^{n_y}$ and $\mathcal{Y}_\mathrm{f}\eqdef\range(R_\mathrm{f})\subseteq\R^{n_y}$. Based on the principal angles between $\mathcal{Y}_\mathrm{s}$ and $\mathcal{Y}_\mathrm{f}$, a decoupling matrix is derived that transforms the original system into a set of SISO and a set of two-inputs, single-output (TISO) systems.  But when the subspace generated by the fast actuators is entirely contained in the subspace generated by slow actuators ($\mathcal{Y}_\mathrm{f}\subset\mathcal{Y}_\mathrm{s}$), the analysis of the principal angles becomes redundant and the use of heuristics unavoidable, leaving the decoupling process unspecified.

Based on the assumption that the bandwidths  of $g_\mathrm{s}(s)$ and $g_\mathrm{f}(s)$ differ significantly, other approaches split the control problem~\eqref{eq:system} into two loops: one feedback loop for the slow array that may be operated at a lower sampling/actuation frequency, and a separate feedback loop for the fast array. Such a separation is implemented in most synchrotrons that use a separate sets of slow and fast correctors~\cite{HUBERT,FOFBALS,PLOUVIEZ}, but interactions at intermediate frequencies can require the introduction of a frequency deadband between the slow and fast systems.  Depending on the disturbance spectrum, this approach can lead to significant performance degradation~\cite{FOFBALS}.

One common way to avoid introducing a frequency deadband is to subtract the predicted effect of the slow array from the feedback signal of the fast array~\cite{HUBERT}. Another solution is to periodically subtract the DC gain from each fast actuator (individually) and to import these values into the slow feedback loop, thereby shifting the low-frequency action from the fast actuator array to the slow actuator array~\cite{PERIODICDC}. However, this approach neglects the coupling between slow and fast actuators and relies on a SISO analysis of the combined slow and fast loops. As in the case of the single-array system~\eqref{eq:CDsystem}, large condition numbers $\kappa(R_\mathrm{s})$ and $\kappa(R_\mathrm{f})$ of the order of $10^3$ to $10^4$ significantly limit the performance, and neglecting the coupling may require further reduction of controller gains~\cite{PLOUVIER}. None of these approaches -- the frequency deadband method~\cite{HUBERT}, the periodic DC method~\cite{PERIODICDC}, or combinations of those~\cite{PLOUVIEZOLD,PLOUVIER} -- provide a means of jointly investigating the stability, performance and robustness properties of the combined feedback loops, which might be subject to instabilities due to large $\kappa(R_\mathrm{s})$ and $\kappa(R_\mathrm{f})$.

In this paper we propose a design method based on the \emph{generalised singular value decomposition} (GSVD) \cite[Ch.\ 6.1.6]{GOLUB4} to decouple the system in~\eqref{eq:system} into sets of TISO and SISO systems. The GSVD factors the matrices $R_\mathrm{s}$ and $R_\mathrm{f}$ as $R_\mathrm{s}=X\Sigma_\mathrm{s} \trans{U}_\mathrm{s}$ and $R_\mathrm{f}=X\Sigma_\mathrm{f} \trans{U}_\mathrm{f}$, where $X$ is invertible, $U_\mathrm{s}$ and $U_\mathrm{f}$ are orthogonal and $\Sigma_\mathrm{s}$ and $\Sigma_\mathrm{f}$ are diagonal and possibly padded with zeroes (Theorem~\ref{thm:gsvd}). By substituting the GSVD in~\eqref{eq:system}, each response matrix is diagonalised, which we refer to as a \emph{generalised modal transformation} (Section~\ref{sec:mainresults}). We show that the output transformation matrix $X$ is closely related to the hypothetical modal transformation of~\eqref{eq:CDsystem} when $R=\begin{bmatrix}R_\mathrm{s} & R_\mathrm{f}\end{bmatrix}$.

In contrast to standard modal decomposition, the mapping to the generalised modal space is defined by the non-orthogonal matrix $X$, which means that the performance properties of the control loop are \emph{not} retained when transforming the decoupled systems back to original space. In particular, if $\mathcal{Y}_\mathrm{f}\subset\mathcal{Y}_\mathrm{s}$, the performance of the fast actuator array, $R_\mathrm{f}g_\mathrm{f}(s)u_\mathrm{f}(s)$, may degrade for certain disturbance directions.  In Section~\ref{sec:pre} we show how the GSVD can be used to define an optimal static compensator for the case that an identical controller is used for each actuator array. Moreover, since any feedback signal is multiplied by $\inv{X}$, an ill-conditioned $X$ leads to large controller gains in disturbance directions aligned with (standard) left singular vectors of $X$ associated with small-magnitude singular values. The resulting control system is prone to instabilities caused by uncertainties in $R_{(\cdot)}$~\cite{ILLCONDPLANTS}.  Analogously to the single-array case~\cite{SANDIRAWINDUP}, we propose a method to balance the controller gains using a regularised inverse of $X$ (Section~\ref{sec:post}).

The paper is organised as follows. Section~\ref{sec:background} summarises the modal decomposition for the single-array case and Section~\ref{sec:mainresults} defines the generalised modal decomposition for the two-array case. Static pre- and post-compensators accounting for the non-orthogonal transformation are introduced in Section~\ref{sec:compensators} and the robustness is analysed in Section~\ref{sec:robustness}. The paper concludes with real-world results from experiments at Diamond in Section~\ref{sec:application}, where the proposed control algorithm is validated in physical experiments on the accelerator.

Standard notation is used throughout the paper. Positive-definite and positive-semidefinite matrices are denoted by $A\succ 0$ and $A\succeq 0$, respectively. The 2-norm (maximum singular value) is denoted by $\twonorm{A}$. The condition number of a possibly rectangular matrix is defined by $\kappa(A)=\twonorm{A}\twonorm{\pinv{A}}$. The variable $s$ denotes the Laplace variable, which is not to be confused with subscript $\mathrm{s}$.
\section{Background: The Modal Decomposition\label{sec:background}}
Standard modal decompositions~\cite{HEATH} decouple the single-array CD system~\eqref{eq:CDsystem} using the thin SVD of $R$, $R=U\Sigma\trans{V}$, where $\inR{U}{n_y}{n_y}$, $\xTx{U}=I$, $\inR{V}{n_y}{n_u}$, $\xTx{V}=I$, and $\Sigma=\diag\left(\sigma_1,\dots,\sigma_{n_y}\right)\succ 0$ assuming that $\rank(R)=n_y \leq n_u$. Substituting the SVD in~\eqref{eq:CDsystem} and defining
\begin{align}\label{eq:modaltransformation}
\hat{y}(s)\eqdef\trans{U}y(s),\,\,\,\,
\hat{u}(s)\eqdef\trans{V}u(s),\,\,\,\,
\hat{d}(s)\eqdef\trans{U}d(s),
\end{align}
yields the modal representation of~\eqref{eq:CDsystem} as
\begin{align}\label{eq:CDdiagonal}
\hat{y}(s)=\Sigma g(s)\hat{u}(s)+ \hat{d}(s).
\end{align}
In modal space, the dynamics are given by a set of decoupled SISO systems. Because the matrices $U$ and $V$ of the modal transformation~\eqref{eq:modaltransformation} are orthonormal, it holds that $\twonorm{\hat{y}(s)}=\twonorm{y(s)}$ and $\twonorm{\hat{u}(s)}=\twonorm{u(s)}$~\cite[Ch.\ 2.3.5]{GOLUB4}, so that stability properties and 2-norm based upper bounds on performance and robustness measures of the control loop are retained when transforming the modal system back to the original space~\cite{CROSSDIR}.

In this paper, we use the \emph{internal model control} (IMC) structure to design the controller. The IMC structure used at Diamond is shown in Fig.~\ref{fig:imc} in original space, where, in the single-array case, $P(s)\eqdef R g(s)$ and $\bar{P}(s)\eqdef\bar{R}\bar{g}(s)$ are the plant and the plant model, $\Delta(s)$ the uncertainty, $Q(s)$ the IMC filter, $\Gamma$ a \textit{regularisation} gain (Section~\ref{sec:singlearrayG}), which is also referred to as output compensator in the following, and $\Upsilon=I$. The IMC filter is designed using plant inversion and is also referred to as a Dahlin or \emph{lambda} controller~\cite[Ch.\ 4.5]{morari1989robust}.
\section{The Generalised Modal Decomposition\label{sec:mainresults}}
To decouple the two-array system~\eqref{eq:system}, the original GSVD formulation~\cite[Ch.\ 6.1.6]{GOLUB4} is transposed and adapted for the spatial response matrices in Theorem~\ref{thm:gsvd}. Throughout the paper, we assume that the slow actuator array spans the output space and no actuator array has redundant components, i.e.\
\begin{align}
&\rank(R_\mathrm{s})=n_\mathrm{s}=n_y,
&\rank(R_\mathrm{f})=n_\mathrm{f}\leq n_y.\tag{Asm. 1}\label{asm:R}
\end{align}
Systems with $\rank(R_{(\cdot)})>n_y$ can be reformulated to satisfy~\eqref{asm:R}, but systems with $\rank(R_{(\cdot)})< n_y$ can be uncontrollable in the sense of~\cite[Def. 6.4]{SKOGESTADMULTI}.
\begin{theorem}[{GSVD~\cite[Thm.\ 6.1.1]{GOLUB4}}]\label{thm:gsvd}
Given $\inR{R_{(\cdot)}}{n_y}{n_{(\cdot)}}$ satisfying~\eqref{asm:R}, the matrices $R_\mathrm{s}$ and $R_\mathrm{f}$ can be jointly factored as
\begin{align}\label{eq:gsvd}
&R_\mathrm{s} = X\begin{bmatrix}\Sigma_\mathrm{s} & 0\\ 0 & I\end{bmatrix} \trans{U}_\mathrm{s},
&R_\mathrm{f} = X\begin{bmatrix}\Sigma_\mathrm{f}\\ 0\end{bmatrix} \trans{U}_\mathrm{f},
\end{align}
where $\inR{X}{n_y}{n_y}$ with $\det(X)\neq 0$ is the matrix of generalised output modes, $\inR{\Sigma_{(\cdot)}}{n_\mathrm{f}}{n_\mathrm{f}}$ with $\Sigma_{(\cdot)}=\diag(\sigma_{{(\cdot)},1},\dots,\sigma_{{(\cdot)},n_\mathrm{f}})\succ 0$ and ${(\cdot)}=\left\lbrace \mathrm{s},\mathrm{f}\right\rbrace$ are the matrices of generalised singular values that satisfy $\sigma_{\mathrm{s},i}^2 + \sigma_{\mathrm{f},i}^2 = 1$, and $\inR{U_{(\cdot)}}{n_{(\cdot)}}{n_{(\cdot)}}$ with $\trans{U_{(\cdot)}}U_{(\cdot)}=I$ are the matrices of generalised input modes.
\end{theorem}
\noindent The GSVD from Thm.~\ref{thm:gsvd} is substituted in~\eqref{eq:system} to obtain
\begin{equation}\label{eq:system_tmp}
\begin{aligned}
y(s) = X\!\begin{bmatrix}\Sigma_\mathrm{s} & 0\\ 0 & I\end{bmatrix} &\trans{U}_\mathrm{s} g_\mathrm{s}(s) u_\mathrm{s}(s)\\&+ X\!\begin{bmatrix}\Sigma_\mathrm{f}\\ 0\end{bmatrix} \trans{U}_\mathrm{f} g_\mathrm{f}(s) u_\mathrm{f}(s) + d(s).
\end{aligned}
\end{equation}
Left-multiplying~\eqref{eq:system_tmp} with $\inv{X}$ and defining
\begin{equation}\label{eq:transformation}
\begin{aligned}
&\tilde{\mb{y}}(s)\eqdef\inv{\mb{X}}\mb{y}(s),
&&\tilde{\mb{u}}_{(\cdot)}(s)\eqdef\trans{\mb{U}}_{(\cdot)} \mb{u}_{(\cdot)}(s),\\
&\tilde{\mb{d}}(s)\eqdef\inv{\mb{X}} \mb{d}(s),
\end{aligned}
\end{equation}
where ${(\cdot)}=\lbrace \mathrm{s},\mathrm{f}\rbrace$, yields the \emph{generalised modal representation} of~\eqref{eq:system}:
\begin{align}\label{eq:modesystem}
\tilde{y}(s)\!=\!\begin{bmatrix}\Sigma_\mathrm{s} & 0\\ 0 & I\end{bmatrix} g_\mathrm{s}(s) \tilde{u}_\mathrm{s}(s)\!+ \!\begin{bmatrix}\Sigma_\mathrm{f}\\ 0\end{bmatrix} g_\mathrm{f}(s) \tilde{u}_\mathrm{f}(s)\!+\!\tilde{d}(s).
\end{align}
Because the matrices $\Sigma_\mathrm{s}$ and $\Sigma_\mathrm{f}$ are diagonal, the MIMO representation~\eqref{eq:system} is decoupled into $n_\mathrm{f}$ TISO systems and $n_\mathrm{s}-n_\mathrm{f}$ SISO systems in~\eqref{eq:modesystem}. The separation between output directions that are affected by TISO systems and those affected by SISO systems is given by
\begin{equation}\label{eq:split}
\begin{aligned}
\mathcal{Y}_{\mathrm{s}\cap\mathrm{f}}&\eqdef \mathcal{Y}_\mathrm{s}\,\cap\,\mathcal{Y}_\mathrm{f}=\spanv{x_1,\dots,x_{n_\mathrm{f}}},\\
\mathcal{Y}_{\mathrm{s}\backslash\mathrm{f}}&\eqdef \mathcal{Y}_\mathrm{s}\,\backslash\,\mathcal{Y}_\mathrm{f}=\spanv{x_{n_\mathrm{f}+1},\dots,x_{n_y}},
\end{aligned}
\end{equation}
where $x_i\in\R^{n_y}$ are the columns of $X$ and $\mathcal{Y}_{\mathrm{s}\cap\mathrm{f}}$ and $\mathcal{Y}_{\mathrm{s}\backslash\mathrm{f}}$ are referred to as TISO and SISO subspace in the following. Note that~\eqref{eq:split} is a consequence of~\eqref{eq:system} that cannot be altered by the choice of decomposition but, compared to an arbitrary factorisation, the output basis provided by Thm.~\ref{thm:gsvd} is closely related to a hypothetical single-array system, which is shown in the following lemma.
\begin{lemma}\label{thm:X}
Consider the factorisation of $R_\mathrm{s}$ and $R_\mathrm{f}$ from Thm.~\ref{thm:gsvd}, and let $\inR{R\eqdef\begin{bmatrix}R_\mathrm{s} & R_\mathrm{f}\end{bmatrix}}{n_y}{(n_\mathrm{s}+n_\mathrm{f})}$. Then, the standard singular values and standard left singular vectors of $R$ equal those of $X$.
\end{lemma}
\begin{proof}
Note that according to Thm.~\ref{thm:gsvd}, the generalised singular values satisfy $\Sigma_\mathrm{s}^2+\Sigma_\mathrm{f}^2=I$. Express $\trans{R}$ using~\eqref{eq:gsvd} and compute $R\trans{R}=X\diag(\Sigma_\mathrm{s}^2+\Sigma_\mathrm{f}^2,\,\, I)\trans{X} = X\trans{X}$, from which the claim follows.
\end{proof}
\noindent Lemma~\ref{thm:X} shows that the decomposition of the two-array system~\eqref{eq:system} through Thm.~\ref{thm:gsvd} relates to the modal decomposition of a hypothetical single-array system~\eqref{eq:CDsystem} with $n_u=n_\mathrm{s}+n_\mathrm{f}$ and $R=\begin{bmatrix}R_\mathrm{s} & R_\mathrm{f}\end{bmatrix}$ and therefore allows the TISO and SISO subspaces~\eqref{eq:split} to be related to the standard left singular vectors of $R$, which determine the spatial distribution of the disturbance spectrum (Section~\ref{sec:disturbance}). Consider the standard SVD $R=U\Sigma\trans{U}$, then $X$ can be formed as $X=U\Sigma\trans{V_X}$, where $V_X$ with $\xTx{V_X}=I$ is the matrix of standard right singular vectors of $X$. The mapping of a vector $y\in\R^{n_y}$ to generalised modal space, $\inv{X}y=V_X\inv{\Sigma}\trans{U}y$, therefore consists of mapping $y$ to mode space first before scaling by $\inv{\Sigma}$ and finally multiplying it with $\trans{V_X}$. The following lemma characterises the gain ratio between each actuator array using a function $f:\mathcal{Y}_{\mathrm{s}\cap\mathrm{f}}\mapsto\R_{\geq 1}$, where $\R_{\geq 1}\eqdef\R\cap\set{x\in\R}{x\geq 1}$, that has been used for a variational formulation of the GSVD~\cite{GSVDVAR}.
\begin{lemma}\label{thm:f}
Consider the function $f:\mathcal{Y}_{\mathrm{s}\cap\mathrm{f}}\mapsto\R_{\geq 1}$, 
\begin{align}
f(y)\eqdef\frac{1}{2}\left(\twonormn{\trans{y}R_\mathrm{s}}^2/\twonormn{\trans{y}R_\mathrm{f}}^2+\twonormn{\trans{y}R_\mathrm{f}}^2/\twonormn{\trans{y}R_\mathrm{s}}^2\right),
\end{align} 
and its gradient $\nabla f:\mathcal{Y}_{\mathrm{s}\cap\mathrm{f}}\mapsto\R^{n_y},$
\begin{equation}\label{eq:gradf}
\begin{aligned}
\nabla f(y) \eqdef&\,\frac{1}{\twonormn{\trans{R_\mathrm{f}}y}^2}\left(R_\mathrm{s}\trans{R_\mathrm{s}}y -\frac{\twonormn{\trans{R_\mathrm{s}}y}^2}{\twonormn{\trans{R_\mathrm{f}}y}^2}R_\mathrm{f}\trans{R_\mathrm{f}}y\right)\,\\&\quad
+\frac{1}{\twonormn{\trans{R_\mathrm{s}}y}^2}\left(R_\mathrm{f}\trans{R_\mathrm{f}}y -\frac{\twonormn{\trans{R_\mathrm{f}}y}^2}{\twonormn{\trans{R_\mathrm{s}}y}^2}R_\mathrm{s}\trans{R_\mathrm{s}}y\right).
\end{aligned}
\end{equation}
It holds that $\nabla f(y)=0$ at $y_i\eqdef\inv{(X\trans{X})} x_i$, where $x_1,\dots,x_{n_y}$ are the columns of $X$ from Thm.~\ref{thm:gsvd}. In addition, $\nabla f(y_i)=0$ and $f(y_i)=1$ if $y_i$ is a shared standard left singular vector for $R_\mathrm{s}$ and $R_\mathrm{f}$ associated with an identical singular value.
\end{lemma}
\begin{proof}
Express $R_{(\cdot)}\trans{R_{(\cdot)}}$ using~\eqref{eq:gsvd}, so that $R_{(\cdot)}\trans{R_{(\cdot)}}y_i=R_{(\cdot)}\trans{R_{(\cdot)}}\inv{(X\trans{X})}x_i=X\Sigma^2_{(\cdot)}\inv{X}x_i=\sigma_{{(\cdot)},i}^2 x_i$, where $\sigma_{{(\cdot)},i}^2$ is a generalised singular value and ${(\cdot)}=\lbrace \mathrm{s},\mathrm{f}\rbrace$. Substituting in the first term on the right-hand side of~\eqref{eq:gradf} yields
\begin{align*}
R_\mathrm{s}\trans{R_\mathrm{s}}y_i -\frac{\twonormn{\trans{R_\mathrm{s}}y_i}^2}{\twonormn{\trans{R_\mathrm{f}}y_i}^2}R_\mathrm{f}\trans{R_\mathrm{f}}y_i=
\sigma_{\mathrm{s},i}^2 x_i - \frac{\sigma_{\mathrm{s},i}^2}{\sigma_{\mathrm{f},i}^2} \sigma_{\mathrm{f},i}^2 x_i = 0,
\end{align*}
and similarly for the second term on the right-hand side of~\eqref{eq:gradf}. It follows that $\nabla f(y_i)=0$. 

For the second part of the claim, note that if $x_i$ is a shared standard left singular vector for $R_\mathrm{s}$ and $R_\mathrm{f}$, then it must be one for $R$ from Lemma~\ref{thm:X} and $X$ as well. It follows that $y_i=x_i/\hat{\sigma}_i^2$, where $\hat{\sigma}_i$ is the corresponding standard singular value of $R$, and hence $\twonormn{\trans{y_i}R_\mathrm{s}}^2=\twonormn{\trans{y_i}R_\mathrm{f}}^2$.
\end{proof}
The function $f(y)$ can be interpreted as a measure for the ratio of gains that are required by each actuator array to produce a correction $y$. In steady state, the actuator effort required to produce a correction is $y=R_\mathrm{s}u_\mathrm{s}+R_\mathrm{f}u_\mathrm{f}$, so that the terms $\twonormn{\trans{y}R_\mathrm{s}}\geq \twonormn{\trans{y}R_\mathrm{s} u_\mathrm{s}}/\twonormn{u_\mathrm{s}}$ and $\twonormn{\trans{y}R_\mathrm{f}}\geq \twonormn{\trans{y}R_\mathrm{f} u_\mathrm{f}}/\twonormn{u_\mathrm{f}}$ can be seen as the relative contribution of each actuator array. If a vector $y_i$ exist that is a shared standard left singular vector associated with standard singular values $\hat{\sigma}_{\mathrm{s},i}$ and $\hat{\sigma}_{\mathrm{f},i}$ for $R_\mathrm{s}$ and $R_\mathrm{f}$, then $f(y_i)=\frac{1}{2}\left(\hat{\sigma}_{\mathrm{s},i}^2/\hat{\sigma}_{\mathrm{f},i}^2+\hat{\sigma}_{\mathrm{f},i}^2/\hat{\sigma}_{\mathrm{s},i}^2\right)$.%
\section{Compensators\label{sec:compensators}}
\begin{figure}
\inputtikz{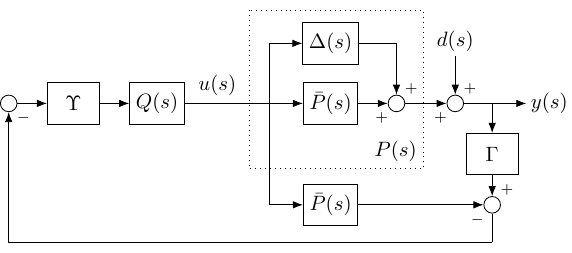}
\caption{Internal model control (IMC) structure with plant $P(s)\eqdef\bar{P}(s)+\Delta(s)$, uncertainty $\Delta(s)$, plant model $\bar{P}(s)$ and static compensators $\Gamma$ and $\Upsilon$.\label{fig:imc}}
\end{figure}

As in the single-array case (Section~\ref{sec:background}), the IMC structure is used but the generalised modal decomposition could also be combined with other feedback structures. The IMC structure is shown in Fig.~\ref{fig:imc}, where, in the two-array case, $u(s)\eqdef\begin{pmatrix}
\trans{u}_\mathrm{s}(s) &\trans{u}_\mathrm{f}(s)\end{pmatrix}^\Tr$, $P(s)\eqdef \begin{bmatrix}R_\mathrm{s}g_\mathrm{s}(s) & R_\mathrm{f}g_\mathrm{f}(s)\end{bmatrix}$, $\bar{P}(s)\eqdef \begin{bmatrix}\bar{R}_\mathrm{s}\bar{g}_\mathrm{s}(s) & \bar{R}_\mathrm{f}\bar{g}_\mathrm{f}(s)\end{bmatrix}$, $Q(s)\eqdef \diag\left(Q_\mathrm{s}(s),\,Q_\mathrm{f}(s)\right)$, and $\Upsilon\eqdef \begin{bmatrix}\trans{\Upsilon}_\mathrm{s}& \trans{\Upsilon}_\mathrm{f}\end{bmatrix}^\Tr$. The matrix $\bar{P}(s)$ is a nominal model of the real plant $P(s)\eqdef\bar{P}(s)+\Delta(s)$ and $\Upsilon$ and $\Gamma$ are (static) input and output compensators introduced in Sections~\ref{sec:pre} and~\ref{sec:post}. As opposed to the standard feedback structure, the main advantage of IMC is that the closed-loop properties are directly related to the open-loop transfer function and not to the inverse of the return difference. Other advantages are that the IMC structure is naturally amenable to plants with time delays~\cite[Ch.\ 3.5]{morari1989robust} and the feedback signal can be used as an input to a fault detection algorithm.

Before considering the input and output compensators, so that $\Gamma=I$ and $\Upsilon=\left[I\quad I\right]^\Tr$, the feedback laws for the decoupled system~\eqref{eq:modesystem} are assumed to be given by
\begin{subequations}\label{eq:inputs}%
\begin{align}%
\tilde{u}_\mathrm{s}(s)&= -\tilde{Q}_\mathrm{s}(s)\tilde{d}(s)\eqdef -\begin{bmatrix}\inv{\Sigma}_\mathrm{s} & 0\\ 0 & I\end{bmatrix} q_{\mathrm{s}}(s) \tilde{d}(s),\\[0.25em]
\tilde{u}_\mathrm{f}(s)&= -\tilde{Q}_\mathrm{f}(s)\tilde{d}(s)\eqdef -\begin{bmatrix}\inv{\Sigma}_\mathrm{f} & \,\,0\end{bmatrix} q_{\mathrm{f}}(s) \tilde{d}(s),
\end{align}%
\end{subequations}
where $q_{\mathrm{s}}:\C\mapsto\C$ and $q_{\mathrm{f}}:\C\mapsto\C$ are stable and realisable transfer functions. The filters are recovered in the original space as $Q_{(\cdot)}(s)=U_{(\cdot)}\tilde{Q}_{(\cdot)}(s)\inv{X}$. Substitute~\eqref{eq:inputs} in~\eqref{eq:modesystem} to obtain the transfer function $\tilde{S}:\C^{n_y}\mapsto\C^{n_y}$ from $\tilde{d}(s)$ to $\tilde{y}(s)$ for $\Delta(s)=0$ as
\begin{equation}\label{eq:CLmode}
\begin{aligned}
\tilde{y}(s) &= \left(I -\Sigma_\mathrm{s}g_\mathrm{s}(s)\tilde{Q}_\mathrm{s}(s)-\begin{bmatrix}\Sigma_\mathrm{f}\\ 0\end{bmatrix}g_\mathrm{f}(s)\tilde{Q}_\mathrm{f}(s)\right)\tilde{d}(s),\\
&=\left(I-I g_\mathrm{s}(s)q_\mathrm{s}(s)-\begin{bmatrix}I & 0\\ 0 & 0\end{bmatrix}g_\mathrm{f}(s)q_\mathrm{f}(s)\right)\tilde{d}(s),\\
&\reqdef\begin{bmatrix}S_{\mathrm{s}\cap\mathrm{f}}(s)I & 0\\ 0 & S_{\mathrm{s}\backslash\mathrm{f}}(s)I\end{bmatrix}\tilde{d}(s),\\
&\reqdef\tilde{S}(s)\tilde{d}(s),
\end{aligned}
\end{equation}
from which it can be seen that the structure~\eqref{eq:inputs} results in output sensitivities that are identical for each TISO and each SISO mode, i.e.\
\begin{equation}\label{asm:S}
\begin{aligned}
S_i(s) &= S_{\mathrm{s}\cap\mathrm{f}}(s),\qquad i=1,\dots,n_\mathrm{f},\\
S_j(s) &= S_{\mathrm{s}\backslash\mathrm{f}}(s),\qquad j=n_\mathrm{f}+1,\dots,n_y,
\end{aligned}
\end{equation}
where $S_i:\C\mapsto\C$ is the transfer function from component $i$ of $\tilde{d}(s)$ to component $i$ of $\tilde{y}(s)$ and $S_{\mathrm{s}\cap\mathrm{f}}:\C\mapsto\C$ and $S_{\mathrm{s}\backslash\mathrm{f}}:\C\mapsto\C$ the TISO and SISO output sensitivities, respectively. Restricting the controller dynamics to a scalar function as in~\eqref{eq:inputs}, is a design constraint commonly accepted for single-array CD control~\cite{SANDIRAOPTIMAL}. For the two-array system, this restriction allows a static (frequency-independent) compensator to be designed that guarantees identical performance in original and generalised modal space (Section~\ref{sec:pre}). To ensure a zero steady-state for disturbances with non-zero offsets, it is assumed that the output sensitivities satisfy
\begin{align}\label{asm:ss}
S_{\mathrm{s}\cap\mathrm{f}}(0)= S_{\mathrm{s}\backslash\mathrm{f}}(0)=0,
\end{align}
which means for $q_\mathrm{s}(s)$ and $q_\mathrm{f}(s)$ that $g_\mathrm{s}(0)q_\mathrm{s}(0)=1$ and $g_\mathrm{f}(0)q_\mathrm{f}(0)=0$.

Inverting the transformations~\eqref{eq:transformation} to map the output sensitivity back to the original space gives:
\begin{equation}\label{eq:CLorig}
\begin{aligned}
S(s) \eqdef&\,\,X \tilde{S}(s)\inv{X},\\
=&\,\,I-I g_\mathrm{s}(s)q_\mathrm{s}(s)-X\begin{bmatrix}I & 0\\ 0 & 0\end{bmatrix}\inv{X}g_\mathrm{f}(s)q_\mathrm{f}(s).
\end{aligned}
\end{equation}
From~\eqref{eq:CLorig}, it can be seen that if $\rank(R_{(\cdot)})=n_y$, then
\begin{align}\label{eq:twonormorig}
\twonormn{S(s)}=\twonormn{\tilde{S}(s)}=\max\lbrace\abs{S_{\mathrm{s}\cap\mathrm{f}}(s)},\abs{S_{\mathrm{s}\backslash\mathrm{f}}(s)}\rbrace,
\end{align}
but if $\rank(R_\mathrm{f})<\rank(R_\mathrm{s})=n_y$, $\twonormn{S(s)}\neq\twonormn{\tilde{S}(s)}$ and an upper bound on $\twonormn{S(s)}$ is given by
\begin{align}\label{eq:performancedifference}
\twonormn{S(s)}=\twonormn{X\tilde{S}(s)\inv{X}}\leq\kappa(X)\twonormn{\tilde{S}(s)}.
\end{align}
Hence if $R$ is ill-conditioned, as is the case for synchrotrons, then from Lemma~\ref{thm:X}, $X$ is ill-conditioned and the performance of the control system in original space can be arbitrarily poor. In the following section, the input compensator $\Upsilon$ is designed to remove the potential performance difference highlighted in~\eqref{eq:performancedifference}.
\subsection{Input Compensator\label{sec:pre}}
Consider the output sensitivity in original $S(s)$ space~\eqref{eq:CLorig} and the IMC structure from Fig.~\ref{fig:imc}, where for the remainder of this section it is assumed that $\Gamma=I$, $\Delta(s)=0$, and $\bar{P}(s)=P(s)$. In~\eqref{eq:CLorig}, the matrix in the term associated with $u_\mathrm{f}(s)$ is
\begin{align}\label{eq:defXsf}
X\begin{bmatrix}I & 0\\ 0 & 0\end{bmatrix}\inv{X} = \begin{bmatrix}X_{11} & 0\\ X_{21} & 0\end{bmatrix}\inv{X}\reqdef X_{\mathrm{s}\cap\mathrm{f}}\inv{X},
\end{align}
where $\inR{X_{11}}{n_\mathrm{f}}{n_\mathrm{f}}$ and $\inR{X_{21}}{(n_\mathrm{s}-n_\mathrm{f})}{n_\mathrm{f}}$. Since~\eqref{eq:defXsf} is responsible for the potential performance difference, it seems natural to set $\Upsilon_\mathrm{s}=I$ and include $\Upsilon_\mathrm{f}$ in the control law~\eqref{eq:inputs}, i.e.\ \[\tilde{u}_\mathrm{f}(s)= -\begin{bmatrix}\inv{\Sigma}_\mathrm{f} & 0\end{bmatrix}\Upsilon_\mathrm{f} q_\mathrm{f}(s) \tilde{y}(s),\] so that $u_\mathrm{s}(s)$ and $u_\mathrm{f}(s)$ are given in the original space by
\begin{subequations}\label{eq:inputf}
\begin{align}
u_\mathrm{s}(s)&= -U_\mathrm{s}\begin{bmatrix}\inv{\Sigma}_\mathrm{s} & 0\\ 0& I\end{bmatrix}\inv{X} q_\mathrm{s}(s) d(s),\\
u_\mathrm{f}(s)&= -U_\mathrm{f}\begin{bmatrix}\inv{\Sigma}_\mathrm{f} & 0\end{bmatrix}\Upsilon_\mathrm{f}\inv{X} q_\mathrm{f}(s) d(s).
\end{align}
\end{subequations}
To construct $\Upsilon_\mathrm{f}$, the following Lemma~\ref{thm:Xsf} is used.
\begin{lemma}\label{thm:Xsf}
Let $A^\dagger$ be the Moore-Penrose pseudoinverse~\cite[P5.5.2]{GOLUB4} of $\inR{A}{n}{n}$ with $\rank(A)=r\leq n$. Then $AA^\dagger$ is symmetric and has $r$ unity and $n-r$  zero eigenvalues.
\end{lemma}
\begin{proof}
Let the standard SVD of $A$ be given as $A=U\diag(\Sigma,0)\trans{V}$ with $\inR{\Sigma}{r}{r}\succ 0$. Then, $A^\dagger=V\diag(\Sigma^{-1},0)\trans{U}$ and $AA^\dagger=U\diag(I,0)\trans{U}$ with $\inR{I}{r}{r}$.
\end{proof}
\noindent If $\Upsilon_\mathrm{f}$ is chosen as
\begin{align}\label{eq:F}
\Upsilon_\mathrm{f} \eqdef X_{\mathrm{s}\cap\mathrm{f}}^\dagger X ,
\end{align}
the output sensitivity~\eqref{eq:CLorig} becomes
\begin{align}\label{eq:CLorigF}
S(s)=I-I g_\mathrm{s}(s)q_\mathrm{s}(s)-X_{\mathrm{s}\cap\mathrm{f}}X_{\mathrm{s}\cap\mathrm{f}}^\dagger g_\mathrm{f}(s)q_\mathrm{f}(s).
\end{align}
By setting $A=X_{\mathrm{s}\cap\mathrm{f}}$ with $\rank(A)=\rank(X_{\mathrm{s}\cap\mathrm{f}})=n_\mathrm{f}$ in Lemma~\ref{thm:Xsf}, then with $\Upsilon_\mathrm{f}$ defined as in~\eqref{eq:F}, the difference between performance of the original and the generalised modal space vanishes, i.e.\ $\twonormn{S(s)}=\twonormn{\tilde{S}(s)}$. From the structure of $X_{\mathrm{s}\cap\mathrm{f}}$ in~\eqref{eq:defXsf}, the structure of $X_{\mathrm{s}\cap\mathrm{f}}^\dagger$ is
\begin{align*}
X_{\mathrm{s}\cap\mathrm{f}}^\dagger = \begin{bmatrix}Z_{11} & Z_{12}\\ 0 & 0\end{bmatrix},
\end{align*}
where $\inR{Z_{11}}{n_\mathrm{f}}{n_\mathrm{f}}$ and $\inR{Z_{12}}{n_\mathrm{f}}{(n_\mathrm{s}-n_\mathrm{f})}$ must satisfy
\begin{align}\label{eq:charZ}
\begin{bmatrix}Z_{11} & Z_{12}\end{bmatrix}\begin{bmatrix}X_{11} \\ X_{21}\end{bmatrix}=Z_{11}X_{11}+Z_{12}X_{21}=I.
\end{align}
The $n_\mathrm{f}$ output directions from Lemma~\ref{thm:Xsf} that are unaffected by $\Upsilon_\mathrm{f}$ and attenuated by $S_{\mathrm{s}\cap\mathrm{f}}(s)$ therefore lie in $\mathcal{Y}_{\mathrm{s}\cap\mathrm{f}}$, but there exist $n_\mathrm{s}-n_\mathrm{f}$ output directions that are zeroed out by $\Upsilon_\mathrm{f}$ and attenuated by $S_{\mathrm{s}\backslash\mathrm{f}}(s)$.

To see the effect of $\Upsilon_\mathrm{f}$ on the generalised modes, consider mapping~\eqref{eq:CLorigF} to generalised modal space using~\eqref{eq:transformation}:
\begin{align}\label{eq:CLmodeF}
\tilde{S}(s) = I-I g_\mathrm{s}(s)q_\mathrm{s}(s)-\inv{X}X_{\mathrm{s}\cap\mathrm{f}}X_{\mathrm{s}\cap\mathrm{f}}^\dagger X g_\mathrm{f}(s)q_\mathrm{f}(s).
\end{align}
From Lemma~\ref{thm:Xsf}, the 2-norm of~\eqref{eq:CLmodeF} is identical to~\eqref{eq:CLmode}. However, the input compensator has the effect of coupling the TISO modes with the SISO modes. To see this, define $X_{\mathrm{s}\backslash\mathrm{f}}\eqdef X-X_{\mathrm{s}\cap\mathrm{f}}$, and note that by the definition of the pseudo-inverse, the last $n_\mathrm{s}-n_\mathrm{f}$ rows of $\pinv{X_{\mathrm{s}\cap\mathrm{f}}}$ are zero. Then, the term $\inv{X}X_{\mathrm{s}\cap\mathrm{f}}X_{\mathrm{s}\cap\mathrm{f}}^\dagger X$ from~\eqref{eq:CLmodeF} can be expanded as
\begin{equation*}
\begin{aligned}
\inv{X}X_{\mathrm{s}\cap\mathrm{f}}X_{\mathrm{s}\cap\mathrm{f}}^\dagger X 
=\inv{X}X_{\mathrm{s}\cap\mathrm{f}} + \inv{X}X_{\mathrm{s}\cap\mathrm{f}}X_{\mathrm{s}\cap\mathrm{f}}^\dagger X_{\mathrm{s}\backslash\mathrm{f}}
=\begin{bmatrix}I & 0 \\ 0 & 0\end{bmatrix}+\begin{bmatrix}I & 0 \\ 0 & 0\end{bmatrix}\begin{bmatrix}0 & \star \\ 0 & 0\end{bmatrix}
=\begin{bmatrix}I & \star \\ 0 & 0\end{bmatrix},
\end{aligned}
\end{equation*}
where $\star$ is a non-zero block given by $Z_{11}X_{12}+Z_{12}X_{22}$.
%
\subsection{Output Compensator\label{sec:post}}
While the performance difference~\eqref{eq:performancedifference} has been removed by the input compensator, it can be seen from~\eqref{eq:inputf} that the controllers for both arrays are proportional to $\inv{X}$. From Lemma~\ref{thm:X}, $X$ has the same condition number as $R$, so the disturbance directions associated with small-magnitude singular values of $R$ therefore cause large input gains for both actuator arrays, which can lead to actuator saturation. Moreover, if the plant model is inexact, i.e.\ $\bar{P}(s)\neq P(s)$, the resulting control system is likely to be prone to instabilities~\cite{ILLCONDPLANTS}. The output compensator is used to remedy this problem and the following section revisits its design for single-array systems, before modifying it for two-array systems.
\subsubsection{Single-Array Systems\label{sec:singlearrayG}}
Consider the single-array system~\eqref{eq:CDsystem} and the control law in modal space,
\begin{align}\label{eq:uhat}
\hat{u}(s)\eqdef-\hat{Q}(s)\hat{d}(s)\eqdef -\inv{\Sigma}q(s)\hat{d}(s),
\end{align}
where $q:\C\mapsto\C$ is such that $q(s)g(s)=T_\mathrm{m}(s)$ with $T_\mathrm{m}(0)=1$, which results in an overall complementary sensitivity $\hat{T}(s)\eqdef T_\mathrm{m}(s)I$. The standard feedback equivalent of $\hat{Q}(s)$, $\hat{C}(s)\eqdef\invbr{I-\hat{Q}(s)\hat{P}(s)}\hat{Q}(s)$, is given by
\begin{align}\label{eq:Chat}
\hat{C}\left(s\right)= \inv{\Sigma}\frac{q(s)}{1-q(s)g(s)} = \inv{\Sigma}\frac{q(s)}{1-T_\mathrm{m}(s)}.
\end{align}
To accommodate systems with $\kappa(R)=\kappa(\Sigma)\gg 1$, a matrix $\inR{\hat{\Gamma}}{n_y}{n_y}$ is defined as follows:
\begin{align}
\hat{\Gamma}\eqdef \invbr{\Sigma^2+\mu I}\Sigma^2=\diag(\frac{\sigma_1^2}{\sigma_1^2+\mu},\dots,\frac{\sigma_{n_y}^2}{\sigma_{n_y}^2+\mu}),
\end{align}
where the scalar $\mu\geq 0$ is a \emph{regularisation parameter}. Right-multiplying $\hat{C}(s)$ with $\hat{\Gamma}$ modifies the controller as
\begin{align}\label{eq:Chatreg}
\hat{C}(s)\hat{\Gamma}=\invbr{\Sigma^2+\mu I}\Sigma\frac{q(s)}{1-T_\mathrm{m}(s)},
\end{align}
i.e.\ the inverse in~\eqref{eq:Chat} has been replaced with the regularised inverse $\invbr{\Sigma^2+\mu I}\Sigma$, thus attenuating input gains associated with small singular values for $\mu>0$. With the controller defined as in~\eqref{eq:Chatreg}, the diagonal elements of the open loop $\hat{L}(s)\eqdef \Sigma g(s)\hat{C}(s)\hat{\Gamma} \reqdef \diag(\ell_1(s),\dots,\ell_{n_y})$ become
\begin{align}
\ell_i(s)=\frac{\sigma_i^2}{\sigma_i^2+\mu}\,\frac{T_\mathrm{m}(s)}{1-T_\mathrm{m}(s)}.
\end{align}
The regularisation parameter $\mu$ therefore changes the open-loop bandwidth as well as the position of the low-frequency asymptote of the Nyquist diagram of $\ell_i(s)$. Note that for $\sigma_i^2\gg\mu$, the effect of $\mu$ is negligible, whereas for $\sigma_j^2\ll\mu$, the open-loop gain and the closed-loop bandwidth are effectively reduced.
\subsubsection{Two-Array Systems}
\begin{figure}
\inputtikz{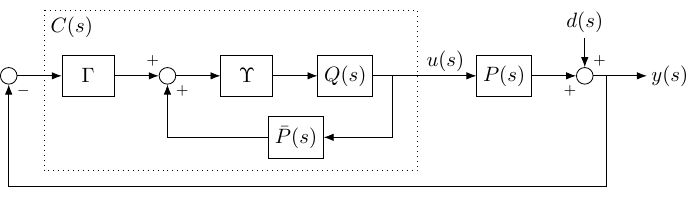}
\caption{IMC structure rearranged into the standard feedback structure for $\Delta(s)=0$.}\label{fig:fb}
\end{figure}
Consider rearranging Fig.~\ref{fig:imc} into the standard feedback structure shown in Fig.~\ref{fig:fb} for $\Delta(s)=0$. For $\Upsilon=\trans{\begin{bmatrix}I & I\end{bmatrix}}$ and $\bar{P}(s)=P(s)$, the controller $C(s)$ obtained as (Appendix)
\begin{equation}\label{eq:standardC}
\begin{aligned}
C(s)\!=\!
U_{\mathrm{s},\mathrm{f}}\!
\begin{bmatrix}\frac{1-S_{\mathrm{s}\backslash\mathrm{f}}(s)}{g_\mathrm{s}(s)S_{\mathrm{s}\cap\mathrm{f}}(s)}\inv{\Sigma_\mathrm{s}} & 0\\[1em] 0 & \frac{1-S_{\mathrm{s}\backslash\mathrm{f}}(s)}{g_\mathrm{s}(s)S_{\mathrm{s}\backslash\mathrm{f}}(s)}I\\[1em] \frac{S_{\mathrm{s}\backslash\mathrm{f}}(s)-S_{\mathrm{s}\cap\mathrm{f}}(s)}{g_\mathrm{f}(s)S_{\mathrm{s}\cap\mathrm{f}}(s)}\inv{\Sigma_\mathrm{f}} & 0\end{bmatrix}
\!\inv{X}\Gamma\!,
\end{aligned}
\end{equation}
where $U_{\mathrm{s},\mathrm{f}}\eqdef\diag(U_{\mathrm{s}},U_{\mathrm{f}})$ and the inverse of $X$ reappears as in~\eqref{eq:inputf}. To attenuate feedback signals that are aligned with directions associated with small singular values of $X$, the inverse $\inv{X}$ in~\eqref{eq:standardC} can be replaced by setting
\begin{align}\label{eq:G:twoarr}
\Gamma\eqdef X\invbr{X^\Tr WX+\mu I}(WX)^\Tr,
\end{align}
for some $\mu>0$ and $W\in\mathcal{S}_{++}$. With $\Gamma$ as defined in~\eqref{eq:G:twoarr}, the term $\inv{X}\Gamma$ in~\eqref{eq:standardC} is replaced by
\begin{align}\label{eq:Xmu}
\inv{X_\mu}\eqdef\invbr{X^\Tr WX+\mu I}(WX)^\Tr,
\end{align}
which can be interpreted as the factor matrix obtained from the following regularised least squares problem~\cite[Ch.\ 6.1.5]{GOLUB4}:
\begin{align*}
\begin{array}{ll}
\displaystyle\minimise_{\nu\in\R^{n_y}} & \twonormn{W^\frac{1}{2}X\nu+b}^2+\mu\twonormn{\nu}^2.
\end{array}
\end{align*}
If $W$ is chosen as $W=U\diag(w_1,\dots,w_{n_y})\trans{U}$, where $U$ is the matrix of standard left singular vectors of $X$, the weights $w_i$ can be chosen to prioritise certain (standard) modes that are particularly affected by disturbances. This follows from Lemma~\ref{thm:X}, which states that $R$ and $X$ share the same matrix of left singular vectors.

Because the matrix $X^\Tr WX$ is positive definite, the 2-norm of $\inv{X_\mu}$ decreases as $\mu$ increases. Let $C_0(s)$ denote the standard controller with $\Gamma=I$ ($\mu=0$). The gain of $C(s)$ for $\Gamma\neq I$ can be upper-bounded by
\begin{equation*}
\twonormn{C(s)} = \twonormn{C_0(s)\inv{X_\mu}}
\leq \twonormn{C_0(s)}\twonormn{\inv{X_\mu}}\leq\twonormn{C_0(s)},
\end{equation*}
from which it can be seen that the parameter $\mu$ controls the bandwidth of the open-loop transfer function $L(s)\eqdef P(s)C(s)$, which is obtained as (Appendix)
\begin{equation}\label{eq:L}
L(s)\!=\!X\!\underbrace{\begin{bmatrix}\frac{1-S_{\mathrm{s}\cap\mathrm{f}}(s)}{S_{\mathrm{s}\cap\mathrm{f}}(s)}I & 0\\
0 & \frac{1-S_{\mathrm{s}\backslash\mathrm{f}}(s)}{S_{\mathrm{s}\backslash\mathrm{f}}(s)}I\end{bmatrix}}_{\reqdef\diag(\ell_{\mathrm{s}\cap\mathrm{f}}(s)I,\,\,\ell_{\mathrm{s}\backslash\mathrm{f}}(s)I)}\!\inv{X}\Gamma\!\reqdef\! L_0(s) \Gamma\!.
\end{equation}
The following lemma formalises the impact of $\Gamma$ on the robustness of the closed-loop system.
\begin{lemma}\label{thm:G}
Suppose that $\rre(\det(L_0(\jw)))>-0.5$ $\forall\omega$ and that the system from Fig.~\ref{fig:imc} is stable for $\Delta(s)=0$. For $\Upsilon=[I\quad I]^\Tr$, the gain margin~\cite[Ch.\ 2.4.5]{SKOGESTADMULTI} of the control system from Fig.~\ref{fig:imc} increases for increasing $\mu$.
\end{lemma}
\begin{proof}By the generalised Nyquist theorem~\cite[Thm. 4.9]{SKOGESTADMULTI}, the closed-loop is stable iff the Nyquist plot of $\det(L(s))$ does not encircle the point $-1+j0$. The claim follows from noting that $\det(L(s))=\det(L_0(s))\det(\Gamma)$ and that $1>\det(\Gamma)>0$ decreases for increasing $\mu$.
\end{proof}
\noindent Note that the compensator can also increase the robustness of systems for which $\rre(\det(L_0(\jw)))\leq-0.5$, which can be verified by comparing $\det(L_0(\jw))$ and $\det(L(\jw))$ in an Argand diagram. For $\Upsilon=[I \quad \Upsilon_\mathrm{f}^\Tr]^\Tr$ with $\Upsilon_\mathrm{f}$ as in~\eqref{eq:F}, the coupling between TISO and SISO modes complicates computing $L_0(s)$, but $L(s)$ remains proportional to $\Gamma$, so that Lemma~\ref{thm:G} remains valid.

\noindent For $\Upsilon=[I \quad \Upsilon_\mathrm{f}^\Tr]^\Tr$ with $\Upsilon_\mathrm{f}$ as in~\eqref{eq:F}, the coupling between TISO and SISO modes complicates computing $L_0(s)$, but $L(s)$ remains proportional to $\Gamma$, so that Lemma~\ref{thm:G} remains valid.

To see how $\Gamma$ affects the closed-loop poles, define $l(s)\eqdef\diag(\ell_{\mathrm{s}\cap\mathrm{f}}(s)I,\,\,\ell_{\mathrm{s}\backslash\mathrm{f}}(s)I)$ and rearrange the complementary sensitivity for $W=I$ as
\begin{gather}\label{eq:muandT}
\begin{aligned}%
T(s) &\,= \inv{(I+L(s))}L(s),\\
&\overset{\eqref{eq:L}}{=}\inv{(I+L_0(s)\Gamma)}L_0(s)\Gamma,\\
&\,=\invbr{I+X l(s)\inv{X_\mu}}X l(s)\inv{X_\mu},\\
&\overset{\eqref{eq:G:twoarr}}{=} X_\mu\invbr{\inv{X}X_\mu+l(s)}l(s)\inv{X_\mu},\\
&\,= X_\mu \invbr{I+\mu\invbr{\xTx{X}}\!\!+l(s)}l(s)\inv{X_\mu}.%
\end{aligned}
\end{gather}
The closed-loop poles are therefore those $s\in\C$ for which
\begin{align}\label{eq:detcondpoles}
\det\left(I+\mu\invbr{\xTx{X}}\!\!+l(s)\right)=0.
\end{align}
For $\mu=0$, the closed-loop poles belong to a subset of $\pi_i(T_{\mathrm{s}\cap\mathrm{f}})\cup\pi_i(T_{\mathrm{s}\backslash\mathrm{f}})\cup\lbrace 0\rbrace$ (Lemma~\ref{thm:G}). To examine the determinant for $\mu>0$, it is further simplified by setting $n_\mathrm{s}\equiv n_\mathrm{f}$, so that with $\ell_{\mathrm{s}\cap\mathrm{f}}(s)=\ell_{\mathrm{s}\backslash\mathrm{f}}(s)=\lambda/s$ \eqref{eq:detcondpoles} becomes:
\begin{equation}\label{eq:detcondpoles2}%
\begin{aligned}%
0&=\det\left(I+\mu\invbr{\xTx{X}}+\frac{\lambda}{s}I\right)\\
&=\det\left(\frac{1}{s}\left(s\left(I+\mu\invbr{\xTx{X}}\right)+\lambda I\right)\right)\\
&=\frac{1}{s^{n_\mathrm{s}}}\det\left(I+\mu\invbr{\xTx{X}}\right)\\
&\qquad\qquad\times\det\left(\lambda\invbr{I+\mu\invbr{\xTx{X}}}+sI\right).
\end{aligned}
\end{equation}
The closed-loop poles are therefore a subset of the eigenvalues of $-\lambda\invbr{I+\mu\invbr{\xTx{X}}}$, which can be obtained by substituting the SVD of $X$:
\begin{align}\label{eq:detcondpoles3}%
\lambda\invbr{I\!+\!\mu\invbr{\xTx{X}}}\!\!= V_X\diag(\lambda \frac{\sigma_i^2}{\sigma_i^2+\mu})\trans{V_X}.
\end{align}
where $\sigma_i$ are the standard singular values of $X$. The poles are therefore $-\lambda\sigma_i^2/(\sigma_i^2+\mu)$, $i=1,\dots,n_y$, and vary from $-\lambda$ for $\mu = 0$ to $0$ for $\mu\rightarrow\infty$ without crossing the real axis (Lemma~\ref{thm:G}). Note that by construction, the root locus of the two-array system with $\ell_{\mathrm{s}\cap\mathrm{f}}(s)=\ell_{\mathrm{s}\backslash\mathrm{f}}(s)=\ell(s)$ corresponds to the root locus of the single-array system.
\begin{remark}
The case $n_{(\cdot)}=n_y$ with $\rank(R_{(\cdot)})=n_y$, ${(\cdot)}=\lbrace\mathrm{s}, \mathrm{f}\rbrace$, considerably simplifies the analysis. In this case, the input compensator $\Upsilon$ becomes redundant, and the open-loop~\eqref{eq:L} simplifies to $L(s)=\ell_{\mathrm{s}\cap\mathrm{f}}(s)X \inv{X_\mu}$. Using the standard SVD of $X$, the matrix $L(s)$ is obtained as
\begin{align}\label{eq:simplecase}
L(s)=\ell_{\mathrm{s}\cap\mathrm{f}}(s)U\diag\left(\frac{\sigma_1^2}{\sigma_1^2+\mu},\dots,\frac{\sigma_{n_y}^2}{\sigma_{n_y}^2+\mu}\right)\trans{U},
\end{align}
where $\sigma_i$ are the standard singular values of $X$. The open loop~\eqref{eq:simplecase} corresponds to the open-loop transfer function of a single-array system designed using the procedure from Section~\ref{sec:singlearrayG}. From Lemma~\ref{thm:X}, the standard singular values equal those of $R=\begin{bmatrix}R_\mathrm{s} & R_\mathrm{f}\end{bmatrix}$. Hence, ignoring model uncertainty, the two-array approach yields the same closed-loop dynamics as a hypothetical single-array system with $n_\mathrm{s}+n_\mathrm{f}$ actuators and $R=\begin{bmatrix}R_\mathrm{s} & R_\mathrm{f}\end{bmatrix}$. Also, considering that $R$ has a kernel of dimension $n_\mathrm{f}$, this argument can be extended to single-array systems with $n_\mathrm{s}$ actuators and $\inR{\bar{R}}{n_\mathrm{s}}{n_\mathrm{s}}$ obtained from the thin SVD of $R$.
\end{remark}
\section{Robust Stability}\label{sec:robustness}
Suppose that the real plant is given by
\begin{equation}
\begin{aligned}
\mb{P}(s)= \bar{\mb{P}}(s)+\begin{bmatrix}\mb{\Delta}_\mathrm{s}g_\mathrm{s}(s) & \mb{\Delta}_\mathrm{f}g_\mathrm{f}(s)\end{bmatrix}
\reqdef\bar{\mb{P}}(s)+\mb{\Delta}(s),
\end{aligned}
\end{equation}
where $\inR{\mb{\Delta}_\mathrm{s}}{n_y}{n_\mathrm{s}}$ and $\inR{\mb{\Delta}_\mathrm{f}}{n_y}{n_\mathrm{f}}$ model the uncertainty for the slow and fast actuator arrays. It is assumed that $g_\mathrm{s}(s)$ and $g_\mathrm{f}(s)$ reflect the actuator dynamics accurately or that any dynamic uncertainties occur at high frequencies at which the controller has a small amplification.

In generalised modal space, the uncertain system is given by
\begin{equation}\label{eq:modesystemdelta}
\begin{aligned}
\tilde{y}(s) = &\left(\begin{bmatrix}\Sigma_\mathrm{s} & 0\\ 0 & I\end{bmatrix}+\tilde{\Delta}_\mathrm{s}\right)g_\mathrm{s}(s) \tilde{u}_\mathrm{s}(s)\\
&\quad+ \left(\begin{bmatrix}\Sigma_\mathrm{f}\\ 0\end{bmatrix}+\tilde{\Delta}_\mathrm{f}\right)g_\mathrm{f}(s) \tilde{u}_\mathrm{f}(s) + \tilde{d}(s),
\end{aligned}
\end{equation}
where $\tilde{\mb{\Delta}}_{(\cdot)}\eqdef\inv{\mb{X}}\mb{\Delta}_{(\cdot)}\mb{U}_{(\cdot)}$. In general, $\tilde{\mb{\Delta}}_{(\cdot)}$ are not diagonal and~\eqref{eq:modesystemdelta} shows that any uncertainty couples the modes in generalised mode space.

The IMC structure with uncertainty is given in Fig.~\ref{fig:imc}. For the robust stability analysis, the transfer function ${\mb{M}}(s)$ from the input to the output of $\Delta(s)$ (see Fig.~\ref{fig:imc}) is
\begin{align}\label{eq:M}
\mb{M}(s) \eqdef - \mb{Q}(s)\Upsilon\invbr{\mb{I}+(\Gamma-\mb{I})\mb{P}(s)\mb{Q}(s)\Upsilon}\Gamma.
\end{align}
The system in Fig.~\ref{fig:imc} is stable iff~\cite[Thm.\ 8.1]{SKOGESTADMULTI}
\begin{align}\label{eq:detcond}
\det \left(\mb{I}-\mb{M}(\mathrm{j}\omega)\mb{\Delta}(\mathrm{j}\omega)\right)\neq 0 \quad\forall\omega.
\end{align}
A sufficient condition for~\eqref{eq:detcond} is
\begin{align}\label{eq:rhocond}
\rho\left(\mb{M}(\mathrm{j}\omega)\mb{\Delta}(\mathrm{j}\omega)\right)=\rho\left(\mb{\Delta}(\mathrm{j}\omega)\mb{M}(\mathrm{j}\omega)\right)
<1\quad \forall\omega,
\end{align}
where $\rho(\cdot)$ denotes the spectral radius, i.e.\ the maximum modulus of the eigenvalues. The product $\mb{\Delta}(s)\mb{M}(s)$ can be rearranged as
\begin{align}
\mb{\Delta}(s)\mb{M}(s)=\underbrace{\begin{bmatrix}\mb{\Delta}_\mathrm{s} & \mb{\Delta}_\mathrm{f}\end{bmatrix}}_{\reqdef\bar{\Delta}}\begin{bmatrix}\mb{I}g_\mathrm{s}(s) & 0\\ 0 & \mb{I}g_\mathrm{f}(s)\end{bmatrix}\mb{M}(s).
\end{align}
For any square matrix $\mb{A}$, the spectral radius can be upper bounded by $\rho(\mb{A})\leq\twonorm{\mb{A}}$~\cite[Ch.\ 7.1.6]{GOLUB4}. A sufficient condition for robust stability is therefore
\begin{align}\label{eq:normcond}
\twonormLR{\bar{\Delta}}
\twonormLR{\begin{bmatrix}\mb{I}g_\mathrm{s}(\mathrm{j}\omega) & 0\\ 0 & \mb{I}g_\mathrm{f}(\mathrm{j}\omega)\end{bmatrix}\mb{M}(\mathrm{j}\omega)}
\!\!<1\quad \forall\omega,
\end{align}
from which an upper bound $\twonormLR{\bar{\Delta}} < \Delta_\text{max}$ on the admissible uncertainty is obtained, where
\begin{align}\label{eq:normdelta}
\Delta_\text{max}\eqdef\min_{\omega\in\R_+}
\twonormLR{\begin{bmatrix}\mb{I}g_\mathrm{s}(\mathrm{j}\omega) & 0\\ 0 & \mb{I}g_\mathrm{f}(\mathrm{j}\omega)\end{bmatrix}\mb{M}(\mathrm{j}\omega)}^{-1}.
\end{align}
Whenever $\twonormLR{\bar{\Delta}}>\Delta_\text{max}$, the control system of Fig.~\ref{fig:imc} \emph{can} be unstable. The right-hand side of~\eqref{eq:normdelta} can be plotted against frequency to find the smallest $\Delta_\text{max}$.
%
%
\section{Results from Diamond Light Source\label{sec:application}}
In view of the Diamond-II upgrade, a more powerful and centralised computing node was integrated in the current Diamond storage ring, aligned with the technical setup of Diamond-II. The upgraded computing node, a Vadatech AMC540~\cite{AMC540}, consists of a Xilinx Virtex-7 FPGA for signal routing and two Texas Instruments C6678 digital signal processors (DSPs) for computing control inputs. This enhancement enables the evaluation of new control algorithms, such as the GSVD-based control approach proposed in this paper.

As the Diamond control problem corresponds to a single-array system~\eqref{eq:CDsystem}, a subset of inputs and outputs is selected and represented as~\eqref{eq:system} with $g_\mathrm{s}(s)= g_\mathrm{f}(s)= g(s)$. The two-array system is decoupled using the GSVD, and the control effort is distributed onto designated slow and fast actuators using \emph{mid-ranging} control~\cite{ALLISON1998469}, which aligns with the configuration of Diamond-II. Both the Diamond and Diamond-II response matrices are ill-conditioned and are therefore comparable. The performance of the two-array controller is then compared to that of a single-array controller.
\subsection{Single-Array Controller\label{sec:DIctrdesign}}
At Diamond, the single-array system~\eqref{eq:CDsystem} has $n_y=173$ BPMs and $n_u=172$ identical magnets. However, the storage ring can be reconfigured, allowing any combination of $n_y\leq 173$ and $n_u\leq 172$ outputs and inputs. The actuator model is~\cite{SANDIRAWINDUP}
\begin{align}\label{eq:DLSI:g}
g(s)\eqdef a/(s+a)\mathrm{e}^{-\tau_d s},
\end{align}
where $a\eqdef 2\pi\times\SI{700}{\radian\per\second}$ and $\tau_d\eqdef\SI{900}{\micro\second}$. For the dynamic part of the controller, the output sensitivity $S_\mathrm{m}:\C\mapsto\C$ is chosen to be
\begin{align}\label{eq:DLSI:S}
S_\mathrm{m}(s) \eqdef 1 - \lambda/(s+\lambda)\mathrm{e}^{-\tau_d s} \reqdef 1-T_\mathrm{m}(s),
\end{align}
where $\lambda\eqdef 1/\tau_d=2\pi\times\SI{176}{\radian\per\second}$ is the closed-loop bandwidth, so that the IMC filter $q:\C\mapsto\C$ is given by 
\begin{align}\label{eq:DLSI:q}
q(s) \eqdef  T_\mathrm{m}(s)/g(s) =  \lambda(s+a)/\left(a(s+\lambda)\right).
\end{align}
To account for the large condition number of $R$, the controller is extended with an output compensator $\Gamma=\inv{(\xTx{R}+\mu I)}\trans{R}$, resulting in the overall control law $u(s)=-K c(s)y(s)$, where $K\eqdef V\diag(k_1,\dots,k_{n_y})\trans{U}$ is the gain matrix, $k_i\eqdef \sigma_i/(\mu+\sigma_i^2)$, and $c(s)$ is given by 
\begin{align}\label{eq:cDI}
c(s)\eqdef \lambda \left(s+a\right)/\left(a\left(s+\lambda\left(1-\mathrm{e}^{-s\tau_d}\right)\right)\right).
\end{align}
In practice, the continuous-time transfer function~\eqref{eq:cDI} is mapped to discrete-time using zero-order hold and implemented as a control law in the form $u_t=-K c(\inv{z})y_t$, where $\inv{z}$ is interpreted as the backwards shift operator and $t\in\mathcal{Z}_+$ represents the discrete-time variable. 

\begin{figure}%
\inputtikz{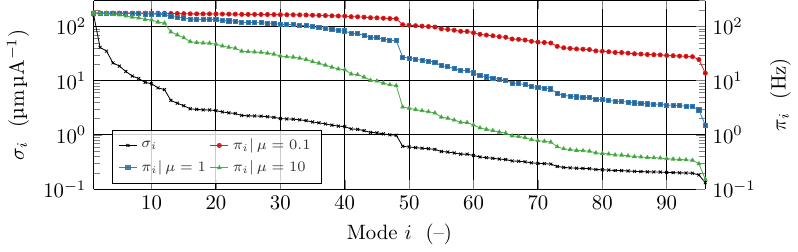}%
\caption{Standard singular values $\sigma_i$ and closed-loop poles $\pi_i$.\label{fig:svs}}%
\end{figure}

The controller~\eqref{eq:cDI} has a pole at $s=0$ and the inclusion of $\Gamma$ effectively reduces the closed-loop bandwidth, which can be seen by computing the overall complementary sensitivity:
\begin{align}\label{eq:modebymodeT}
T(s) = U\!\diag\left(\!\frac{\sigma_1 k_1 \lambda \mathrm{e}^{-s\tau_d}}{s+\tilde{\pi}_1(s)},\,\dots,\frac{\sigma_{n_y} k_{n_y} \lambda \mathrm{e}^{-s\tau_d}}{s+\tilde{\pi}_{n_y}(s)}\!\right)\!\trans{U},
\end{align}
where $\tilde{\pi}_i(s)\eqdef\lambda\left(1-(1-\sigma_i k_i)\mathrm{e}^{-s\tau_d}\right)$. Fig.~\ref{fig:svs} compares the standard singular values $\sigma_i$ (\addlegendimageintext{mark=x, draw=black,fill=black}, left-hand side axis), $i=1,\dots,n_y$, with the corresponding closed-loop poles $\pi_i$ for $\mu\in\lbrace 0.1,1,10\rbrace$ (right-hand side axis). For modes associated with $\sigma_i\gg \mu$, $\pi_i\approx\lambda=2\pi\times\SI{176}{\radian\per\second}$, but as the singular values decrease, the closed-loop bandwidth is reduced, hence increasing the robustness for modes associated with small singular values. At Diamond, the choice $\mu=1$ has proven to be effective. The controller structure would allow different values of $\mu$ to be chosen for different modes, but previous research has shown that $\mu=1$ is near-optimal with respect to some robust performance criterion~\cite{SANDIRAOPTIMAL}.
%
\subsection{Two-Array Controller\label{sec:twoarray:DIIctrdesign}}
To replicate the Diamond-II configuration, $n_\mathrm{s}=96$ magnets are chosen to represent the slow actuators, while $n_\mathrm{f}=64$ are chosen to represent the fast actuators. These $n_u=n_\mathrm{s}+n_\mathrm{f}=160$ magnets are controlling a subset of $n_y=96$ BPMs. The selection of magnets and BPMs is based on physical arguments and on~\eqref{asm:R}, which is applicable to Diamond-II. The matrices $R_\mathrm{s}$ and $R_\mathrm{f}$ are obtained from extracting the corresponding rows and columns of $R$.
\subsubsection{Mid-Ranging Control}
%
%
For the two-array controller, the output sensitivity of the TISO systems is chosen to be
\begin{align}\label{eq:DLSI:Sscapf}
S_{\mathrm{s}\cap\mathrm{f}}(s) \eqdef 1 - \lambda_{\mathrm{s}\cap\mathrm{f}}/(s+\lambda_{\mathrm{s}\cap\mathrm{f}})\mathrm{e}^{-\tau_d s}\reqdef 1-T_{\mathrm{s}\cap\mathrm{f}}(s),
\end{align}
where for later comparison, $\lambda_{\mathrm{s}\cap\mathrm{f}}\eqdef \lambda$ matches the closed-loop bandwidth of the existing single-array design~\eqref{eq:DLSI:S}. The output sensitivity of the SISO systems is chosen to be
\begin{align}\label{eq:DLSI:Ssslashf}
S_{\mathrm{s}\backslash\mathrm{f}}(s) \eqdef 1 - \lambda_{\mathrm{s}\backslash\mathrm{f}}/(s+\lambda_{\mathrm{s}\backslash\mathrm{f}})\mathrm{e}^{-\tau_d s}\reqdef 1-T_{\mathrm{s}\backslash\mathrm{f}}(s),
\end{align}
where some of the following experiments use $\lambda_{\mathrm{s}\backslash\mathrm{f}}\eqdef 2\pi\times\SI{50}{\radian\per\second}$ and some $\lambda_{\mathrm{s}\backslash\mathrm{f}}\eqdef 2\pi\times\SI{10}{\radian\per\second}$. With $S_{\mathrm{s}\cap\mathrm{f}}(s)$ and $S_{\mathrm{s}\backslash\mathrm{f}}(s)$ fixed, the IMC filter for the $u_\mathrm{s}(s)$ array is
\begin{align}\label{eq:DLSI:qs}
q_\mathrm{s}(s)\eqdef T_{\mathrm{s}\backslash\mathrm{f}}(s)/g_\mathrm{s}(s)=\left(\lambda_{\mathrm{s}\backslash\mathrm{f}}(s\!+\!a)\right)/\left(a(s\!+\!\lambda_{\mathrm{s}\backslash\mathrm{f}})\right),
\end{align}
and the filter for the $u_\mathrm{f}(s)$ array is
\begin{equation}\label{eq:DLSI:qf}
\begin{aligned}
q_\mathrm{f}(s)\eqdef&\,\, \left(T_{\mathrm{s}\cap\mathrm{f}}(s)-T_{\mathrm{s}\backslash\mathrm{f}}(s)\right)/g_\mathrm{f}(s),\\
=&\,\,
\frac{\lambda_{\mathrm{s}\cap\mathrm{f}}-\lambda_{\mathrm{s}\backslash\mathrm{f}}}{a}\frac{s(s+a)}{(s+\lambda_{\mathrm{s}\cap\mathrm{f}})(s+\lambda_{\mathrm{s}\backslash\mathrm{f}})}.
\end{aligned}
\end{equation}
The choices~\eqref{eq:DLSI:Sscapf}--\eqref{eq:DLSI:qf} are also referred to as \emph{mid-ranging} controllers~\cite{ALLISON1998469}. The overall bandwidth is split between the slow and fast actuator arrays, which will allow for a higher closed-loop bandwidth at Diamond-II.

\begin{figure}
\inputtikz{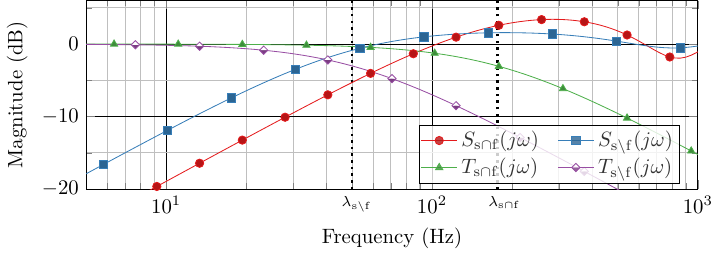}
\caption{Bode plots of $S_{(\cdot)}(s)$ and $T_{(\cdot)}(s)$ for the TISO ($\lambda_{\mathrm{s}\cap\mathrm{f}}=2\pi\times\SI{176}{\radian\per\second}$) and SISO ($\lambda_{\mathrm{s}\backslash\mathrm{f}}= 2\pi\times\SI{50}{\radian\per\second}$) systems.}\label{fig:bodeDIsensitivity}
\end{figure}

Fig.~\ref{fig:bodeDIsensitivity} shows the Bode magnitude plots of the output sensitivities, $S_{\mathrm{s}\cap\mathrm{f}}(\jw)$ (\addlegendimageintext{mark=square*, draw=legbluedraw,fill=legbluefill}) and $S_{\mathrm{s}\backslash\mathrm{f}}(\jw)$ (\addlegendimageintext{mark=*, draw=legreddraw,fill=legredfill}), and the corresponding complementary sensitivities, $T_{\mathrm{s}\cap\mathrm{f}}(\jw)$ (\addlegendimageintext{mark=triangle*, draw=leggreendraw,fill=leggreenfill}) and $T_{\mathrm{s}\backslash\mathrm{f}}(\jw)$ (\addlegendimageintext{mark=halfdiamond*, draw=legvioletdraw,fill=legvioletfill}) for $\lambda_{\mathrm{s}\backslash\mathrm{f}}= 2\pi\times\SI{50}{\radian\per\second}$. Due to the large time delay, the phase lag of $T_{\mathrm{s}\cap\mathrm{f}}(\jw)$ reaches \SI{60}{\degree} at \SI{100}{\Hz}, which significantly reduces the bandwidth of $S_{\mathrm{s}\cap\mathrm{f}}(\jw)$. The sensitivity peaks are $\infnorm{S_{\mathrm{s}\cap\mathrm{f}}(\jw)}=$ \SI{3.4}{\dB} at \SI{290}{\Hz} and $\infnorm{S_{\mathrm{s}\backslash\mathrm{f}}(\jw)}=\SI{0.4}{\dB}$ at \SI{113}{\Hz}.
\subsubsection{Input and Output Compensators}
\begin{figure*}
\centering
\begin{subfigure}[t]{0.46\textwidth}%
	\inputtikz{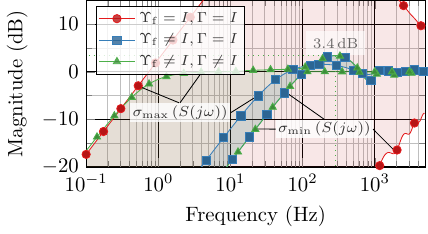}%
	\caption{$S(s)$ for $\lambda_{\mathrm{s}\backslash\mathrm{f}}= 2\pi\times\SI{50}{\radian\per\second}$.\label{fig:bodeDIcompensator:50}}
\end{subfigure}%
\hfill
\begin{subfigure}[t]{0.46\textwidth}%
	\inputtikz{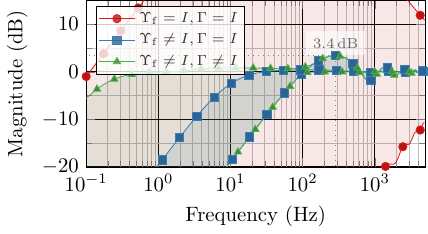}%
	\caption{$S(s)$ for $\lambda_{\mathrm{s}\backslash\mathrm{f}}= 2\pi\times\SI{10}{\radian\per\second}$.\label{fig:bodeDIcompensator:10}}
\end{subfigure}\\[1.5em]%
\begin{subfigure}[t]{0.46\textwidth}%
	\inputtikz{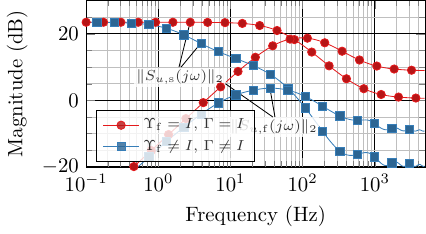}%
	\caption{$S_{u,(\cdot)}(s)$ for $\lambda_{\mathrm{s}\backslash\mathrm{f}}= 2\pi\times\SI{50}{\radian\per\second}$.\label{fig:bodeDIcompensatoracc:50}}
\end{subfigure}%
\hfill
\begin{subfigure}[t]{0.46\textwidth}%
	\inputtikz{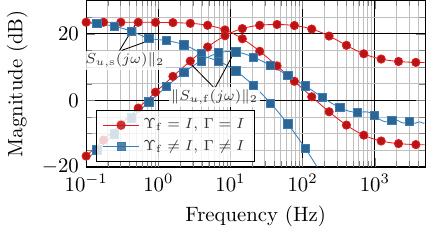}%
	\caption{$S_{u,(\cdot)}(s)$ for $\lambda_{\mathrm{s}\backslash\mathrm{f}}= 2\pi\times\SI{10}{\radian\per\second}$.\label{fig:bodeDIcompensatoracc:10}}
\end{subfigure}
\caption{Minimum and maximum singular values of the sensitivity ($S(s)$) and the transfer functions from $d(s)$ to $u_\mathrm{s}(s)$ and $u_\mathrm{f}(s)$ ($S_{u,\mathrm{s}}(s)$ and $S_{u,\mathrm{f}}(s)$) for $\lambda_{\mathrm{s}\cap\mathrm{f}}=2\pi\times\SI{176}{\radian\per\second}$, $\lambda_{\mathrm{s}\backslash\mathrm{f}}= 2\pi\times\SI{50}{\radian\per\second}$ (a and c), $\lambda_{\mathrm{s}\backslash\mathrm{f}}= 2\pi\times\SI{10}{\radian\per\second}$ (b and d), and different compensators.}\label{fig:bodeDIcompensator}
\end{figure*}
%
While the overall condition number for the two-array system is $\kappa(R)=\kappa(X)=1159$, the corresponding condition numbers for the generalised singular values are $\kappa(\Sigma_\mathrm{s})=4.3$ and $\kappa(\Sigma_\mathrm{f})=4.5$, and therefore allow $\Sigma_\mathrm{s}$ and $\Sigma_\mathrm{f}$ in~\eqref{eq:inputs} to be inverted.

Fig.~\ref{fig:bodeDIcompensator:50} shows the minimum and maximum gains of the output sensitivity~\eqref{eq:CLorig} (\addlegendimageintext{mark=*, draw=legreddraw,fill=legredfill}), $\sigma_\text{min}(S(\jw))$ and $\sigma_\text{max}(S(\jw))$, for $\lambda_{\mathrm{s}\backslash\mathrm{f}}= 2\pi\times\SI{50}{\radian\per\second}$ and for the case that the scalar filters from Section~\ref{sec:twoarray:DIIctrdesign} are embedded in the MIMO system without input and output compensators ($\Upsilon=\trans{\begin{bmatrix}I & I\end{bmatrix}}$ and $\Gamma=I$). For orthogonal $X$, the magnitude of the sensitivity would be enclosed by the TISO and SISO transfer functions from Fig.~\ref{fig:bodeDIsensitivity}, but with the ill-conditioned $X$ and $n_\mathrm{f}<n_\mathrm{s}$, some disturbance directions from \SI{1}{\Hz} to \SI{5}{\kilo\Hz}. The input compensator $\Upsilon=\begin{bmatrix}I & \Upsilon_\mathrm{f}^\Tr\end{bmatrix}^\Tr$ from Section~\ref{sec:pre} removes the performance difference between generalised modal space and original space, so that the ``input-compensated'' sensitivity (\addlegendimageintext{mark=square*, draw=legbluedraw,fill=legbluefill}, Fig.~\ref{fig:bodeDIcompensator:50}) equals the sensitivity from Fig.~\ref{fig:bodeDIsensitivity}. The output compensator $\Gamma$ is obtained from~\eqref{eq:G:twoarr} with $W=I$ and $\mu=1$. Analogous to the single-array case, the output compensator reduces the closed-loop bandwidth, which can be seen from the ``input- and output-compensated'' sensitivity (\addlegendimageintext{mark=triangle*, draw=leggreendraw,fill=leggreenfill}). 

The analysis is repeated in Fig.~\ref{fig:bodeDIcompensator:10} for $\lambda_{\mathrm{s}\backslash\mathrm{f}}= 2\pi\times\SI{10}{\radian\per\second}$. Compared to Fig.~\ref{fig:bodeDIcompensator:50}, the bandwidth of the maximum sensitivity gain $\sigma_\text{max}(S(\jw))$ is reduced, while the bandwidth of $\sigma_\text{min}(S(\jw))$, which is determined by the TISO systems, remains the same. Fig.~\ref{fig:bodeDIcompensator:10} also shows that for the ``input- and output-compensated'' sensitivity (\addlegendimageintext{mark=triangle*, draw=leggreendraw,fill=leggreenfill}), some disturbance directions are amplified between \SI{1}{\Hz} and \SI{100}{\Hz} with a local peak of roughly \SI{1}{\dB} at \SI{10}{\Hz}, which corresponds to the transition between slow and fast actuators in mid-ranging control. This peak does not appear for the ``uncompensated'' sensitivity (\addlegendimageintext{mark=square*, draw=legbluedraw,fill=legbluefill}) and is less pronounced in Fig.~\ref{fig:bodeDIcompensator:50}, and therefore likely to be associated with the additional phase lag introduced by the slow actuator array for $\lambda_{\mathrm{s}\backslash\mathrm{f}}= 2\pi\times\SI{10}{\radian\per\second}$.

The effect of the regularisation on the inputs is illustrated in Fig.~\ref{fig:bodeDIcompensatoracc:50} and~\ref{fig:bodeDIcompensatoracc:10}, which show the maximum gain of the \emph{input sensitivities} of the slow and fast actuator arrays, $S_{u,\mathrm{s}}(\jw)$ and $S_{u,\mathrm{f}}(\jw)$. Without compensators (\addlegendimageintext{mark=*, draw=legreddraw,fill=legredfill}), the control effort is sustained up to $\SI{100}{\Hz}$. Moreover, reducing the SISO bandwidth, such as in Fig.~\ref{fig:bodeDIcompensator:10} and~\ref{fig:bodeDIcompensatoracc:10}, increases the control action of the fast actuator array, which is a consequence of the mid-ranging approach. With compensators (\addlegendimageintext{mark=square*, draw=legbluedraw,fill=legbluefill}), the controller gain is reduced by \SI{20}{\dB} at $\SI{100}{\Hz}$ for both actuator arrays in Fig.~\ref{fig:bodeDIcompensatoracc:50} and~\ref{fig:bodeDIcompensatoracc:10}. Increasing the regularisation parameter $\mu$ would have the effect of decreasing the gains further.
\subsection{Disturbance Spectrum\label{sec:disturbance}}
For both single- and two-array models, all exogenous effects are lumped into the (output) disturbance $d(s)$. The characteristic disturbance can be split into an input disturbance $d_u(s)$ and an output disturbance $d_y(s)$:
\begin{align}\label{eq:distsplit}
d(s) = R d_u(s) + d_y(s).
\end{align}
The contribution from $d_u(s)$ can be associated with ground vibrations and vibrating machine components, which are transmitted to the corrector magnets by the supporting girders and exhibit structural resonances at particular frequencies~\cite{IANDIAMONDIORBIT}. The contribution from $d_y(s)$ is mainly associated with the operation of the synchrotron.
\begin{figure*}[]
    \centering
    \begin{subfigure}[t]{0.49\textwidth}%
        \inputtikz{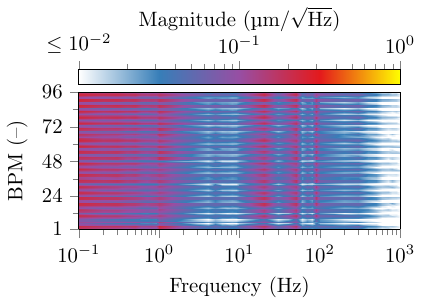}%
        \caption{Original space: min. $=10^{-2.6}$, max. $=10^{-0.6}$ $\si{\micro\meter\per\sqrt{\Hz}}$.\label{fig:psddistorig}}%
    \end{subfigure}%
    \hfill
    \begin{subfigure}[t]{0.49\textwidth}%
        \inputtikz{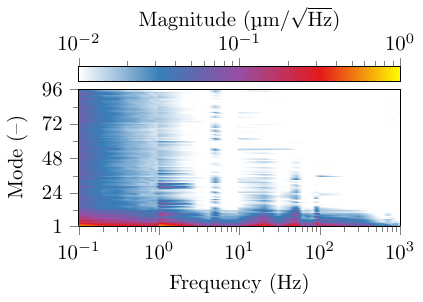}%
        \caption{Mode space: min. $=10^{-2.5}$, max. $=10^{0.1}$ $\si{\micro\meter\per\sqrt{\Hz}}$.\label{fig:psddistmode}}
    \end{subfigure}\\
    \caption{Measured ASD of the disturbance in original and mode space.\label{fig:psddist}}%
\end{figure*}
\begin{figure*}[]
    \centering
    \begin{subfigure}[t]{0.49\textwidth}%
        \inputtikz{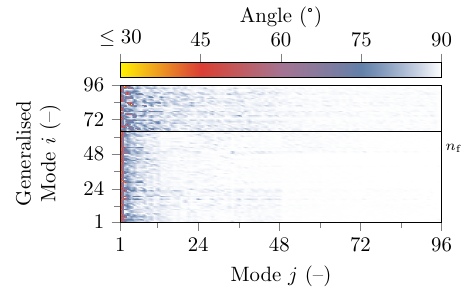}%
        \caption{Angle as $\acos(\abs{\trans{x_i}U_j}/\twonormn{x_i})$: min. = \SI{13}{\degree}, max. = \SI{90}{\degree}.\label{fig:XxUangles}}%
    \end{subfigure}%
    \hfill
    \begin{subfigure}[t]{0.49\textwidth}%
        \inputtikz{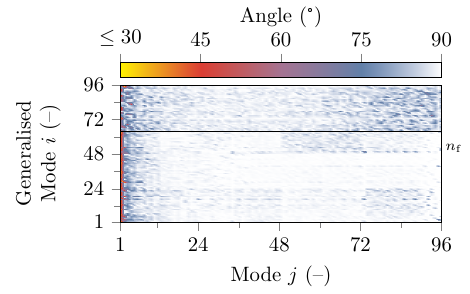}%
        \caption{Angle as $\acos(\abs{\trans{x_{\mu,i}}U_j}/\twonormn{x_{\mu,i}})$ min. = \SI{13}{\degree}, max. = \SI{90}{\degree}.\label{fig:XmuUangles}}
    \end{subfigure}%
    \caption{Angles between columns of $X$ and $U$ (a) and $X_\mu$ and $U$ (b). The horizontal line separates TISO from SISO modes.\label{fig:XUangles}}%
\end{figure*}

The (output) disturbance spectrum can be estimated when the FOFB is disabled, i.e.\ when $y_t=d_t$. Fig.~\ref{fig:psddistorig} shows the \emph{amplitude spectral density} (ASD) for BPMs $i=1,\dots,96$ from \SI{0.1}{\Hz} to \SI{1}{\kilo\Hz}, which is computed from the discrete Fourier transform~\cite[Ch.\ 2.2]{LJUNG} of the measured signal $d_t$ as
\begin{align}\label{eq:ASD}
&D_{(i)}(\omega_k) \eqdef \sqrt{\frac{2}{f_s}}\left\lvert\sum_{t=0}^{N-1}d_{(i),t}\mathrm{e}^{-\frac{\mathrm{j}2\pi kt}{N}}\right\rvert,
\end{align}
where $\omega_k\eqdef 2\pi k f_s/N$, $k=0,\dots,N-1$, $f_s=\SI{10}{\kHz}$ and $N=10^5$, which corresponds to \SI{1}{\second} of data and results in a frequency resolution of $f_s/N=\SI{0.1}{\Hz}$. In practice, the ASD~\eqref{eq:ASD} is computed $10$ times for a signal of length $10N$ and then averaged using Welch's method~\cite[Ch.\ 6.4]{LJUNG}. The horizontal pattern of the ASD in Fig.~\ref{fig:psddistorig} can be associated with girder resonances~\cite{IANDIAMONDIORBIT}, such as the peaks at \SI{0.2}{\Hz}, \SI{1}{\Hz}, \SI{20}{\Hz} and \SI{120}{\Hz}. The vertical pattern is associated with the partitioning of the storage ring, which possesses an (approximate) 24-fold circulant symmetry~\cite{SANDIRAOPTIMAL}. Within each cell, the placement of BPMs on the girders and the distance to other devices determine the sensitivity to disturbances; some BPMs, such as those located downstream of an insertion device, are exposed to larger disturbances than other BPMs.

The disturbance can be mapped to mode space using the orthonormal matrix~\eqref{eq:modaltransformation}. The resulting amplitude spectral density, $\hat{D}_i(\omega)$, is shown in Fig.~\ref{fig:psddistmode}, where the vertical axis refers to the $i$th mode with $i=1$ being associated with the largest (standard) singular value of $R$. Due to the orthonormal property of the transformation matrix $U$, it holds that $\twonormn{D(\omega_k)}^2=\twonormn{\hat{D}(\omega_k)}^2\quad\forall\omega_k$, where the square is applied element-wise. Compared to Fig.~\ref{fig:psddistorig}, the circulant pattern has been attenuated and the ASD is concentrated in the low-order modes ($i\leq 10$) instead. The concentration of the disturbance in the low-order modes justifies the output compensator from Section~\ref{sec:singlearrayG}, which significantly reduces the bandwidth for higher-order modes.

For the two-array case, the ill-conditioned $X$ makes analysing the effect of the characteristic disturbance spectrum onto the performance more difficult. Mapping the disturbance through~\eqref{eq:transformation} would show that the disturbance is spread onto TISO as well as SISO modes. Alternatively, consider computing the acute angles between $x_i$, the columns of $X$, and $U_j$, the standard left singular vectors of $R$, for $i,j=1,\dots,n_y$. An angle equal to \SI{90}{\degree} means that standard mode $j$ does \emph{not} contribute to generalised mode $i$, whereas \SI{0}{\degree} means mode $j$ is parallel to generalised mode $i$, but its particular weight depends on $X$. The angles are computed as $\acos(\abs{\trans{x_i}U_j}/\twonormn{x_i})$ and shown in Fig.~\ref{fig:XxUangles}, where the horizontal axis refers to the $i$th standard mode and the vertical axis to the $j$th generalised mode. Fig.~\ref{fig:XxUangles} shows that most generalised modes form an angle of less than \SI{30}{\degree} with the first mode, i.e.\ the vectors $x_i$ are arranged as a cone of (linearly independent) vectors that is ``centred'' around $U_1$. In fact, the first standard left singular vectors of $R_\mathrm{s}$ and $R_\mathrm{f}$ form an angle of \SI{0.5}{\degree} ($\twonormn{R_\mathrm{s}}=103$ and $\twonormn{R_\mathrm{f}}=112$). Fig.~\ref{fig:XxUangles} also shows that the higher-order standard modes $i=48,\dots,96$ are almost orthogonal to the generalised modes; disturbances aligned to these standard modes require larger gains from all generalised modes when multiplied by $\inv{X}$, which is readily explained through Lemma~\ref{thm:X}. The analysis is repeated in Fig.~\ref{fig:XmuUangles}, which shows the angles between the columns of $X_\mu$~\eqref{eq:Xmu}, i.e.\ the ``regularised basis'', and $U_j$. Compared to Fig.~\ref{fig:XxUangles}, the higher-order standard modes form smaller angles with the generalised modes, whereas the angles for the lower-order modes remain unchanged. With the angles from Fig.~\ref{fig:XmuUangles}, disturbances aligned to higher-order modes will produce smaller gains when multiplied by $\inv{X_\mu}$.
\begin{figure*}
{
\centering
\includegraphics[scale=1]{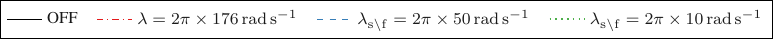}\\[1em]
}
\inputtikz{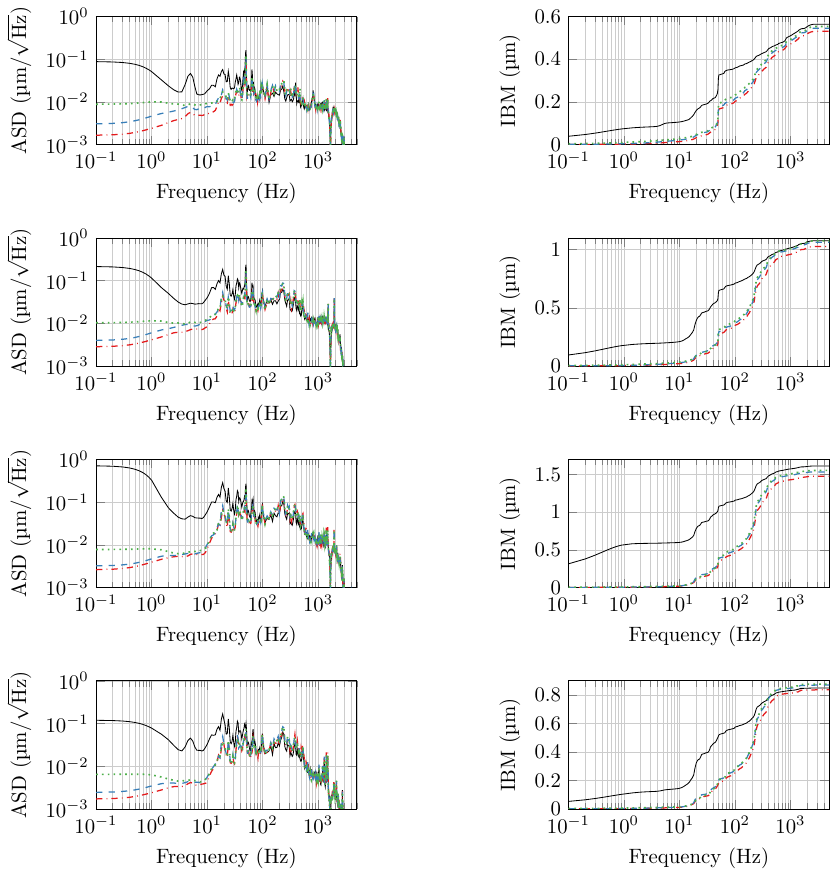}%
\caption{Measured output ASD (left) and IBM (right) for BPMs 1, 3, 5 and 7 of the Diamond storage ring for disabled feedback (OFF), single-array controller ($\lambda$) from Section~\ref{sec:DIctrdesign} and two differently tuned two-array controllers ($\lambda_{\mathrm{s}\backslash\mathrm{f}}$) from Section~\ref{sec:twoarray:DIIctrdesign}.\label{fig:psdy_cell1_XY}}
\end{figure*}
\begin{figure*}[]
    \centering
    \begin{subfigure}[t]{0.45\textwidth}%
        \inputtikz{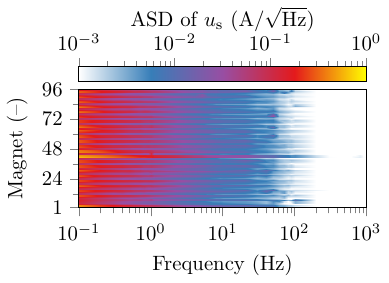}%
        \caption{Min.$=10^{-4.1}$, max.$=10^{-0.2}$ $\si{\ampere\per\sqrt{\Hz}}$.\label{fig:psd_input_gsvd_slow_v105_13112022_Y}}%
    \end{subfigure}
    \hfill
    \begin{subfigure}[t]{0.45\textwidth}%
        \inputtikz{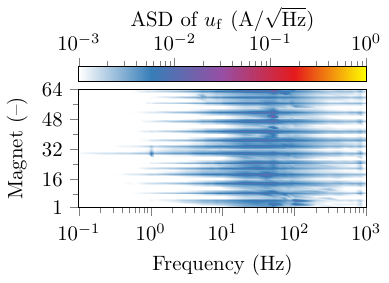}%
        \caption{Min.$=10^{-4.0}$, max.$=10^{-1.7}$ $\si{\ampere\per\sqrt{\Hz}}$.\label{fig:psd_input_gsvd_fast_v105_13112022_Y}}%
    \end{subfigure}\\
    \begin{subfigure}[t]{0.45\textwidth}%
        \inputtikz{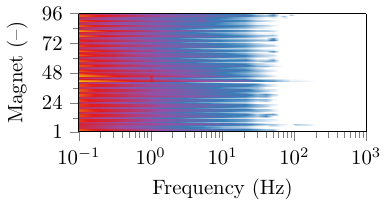}%
        \caption{Min.$=10^{-4.7}$, max.$=10^{-0.2}$ $\si{\ampere\per\sqrt{\Hz}}$.\label{fig:psd_input_gsvd_slow_v106_13112022_Y}}%
    \end{subfigure}
    \hfill
    \begin{subfigure}[t]{0.45\textwidth}%
        \inputtikz{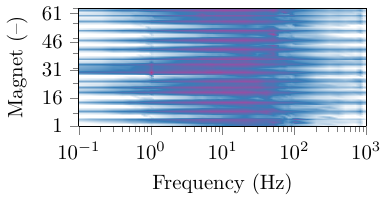}%
        \caption{Min.$=10^{-3.4}$, max.$=10^{-1.3}$ $\si{\ampere\per\sqrt{\Hz}}$.\label{fig:psd_input_gsvd_fast_v106_13112022_Y}}%
    \end{subfigure}\\
    \begin{subfigure}[t]{0.45\textwidth}%
        \inputtikz{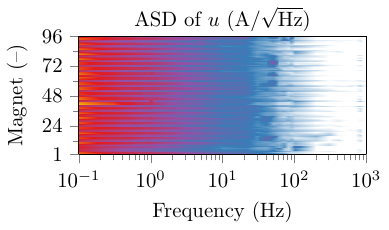}%
        \caption{Min.$=10^{-3.6}$, max.$=10^{-0.3}$ $\si{\ampere\per\sqrt{\Hz}}$.\label{fig:psd_input_imc_v109_13112022_Y}}%
    \end{subfigure}
    \hfill
    \begin{subfigure}[t]{0.45\textwidth}%
    \end{subfigure}
 \caption{Measured ASD of inputs. The first row shows a two-array controller with $\lambda_{\mathrm{s}\backslash\mathrm{f}}=2\pi\times\SI{50}{\radian\per\second}$ and $\lambda_{\mathrm{s}\cap\mathrm{f}}=2\pi\times\SI{176}{\radian\per\second}$, the second row one with $\lambda_{\mathrm{s}\backslash\mathrm{f}}=2\pi\times\SI{10}{\Hz}$ and $\lambda_{\mathrm{s}\cap\mathrm{f}}=2\pi\times\SI{176}{\radian\per\second}$, and the third row a single-array controller with $\lambda_{\mathrm{s}\cap\mathrm{f}}=2\pi\times\SI{176}{\radian\per\second}$.\\\label{fig:psd_input}}
\end{figure*}
\subsection{Results from the Storage Ring\label{sec:results}}
Two versions of the two-array controller -- one with $\lambda_{\mathrm{s}\backslash\mathrm{f}}=2\pi\times\SI{50}{\radian\per\second}$ and one with $\lambda_{\mathrm{s}\backslash\mathrm{f}}=2\pi\times\SI{10}{\radian\per\second}$ -- were tested on the Diamond storage ring. The controllers are implemented in discrete time and parallelised in C language on the DSPs -- one for the horizontal and one for the vertical direction. The FPGA is responsible for signal routing and the combined system is capable of producing control inputs at rates of at least $\SI{10}{\kHz}$.
\subsubsection{Outputs}
The left-hand side of Fig.~\ref{fig:psdy_cell1_XY} shows the output ASD measured in the first cell of the Diamond storage ring for disabled feedback (\addlegendimageintext{mark=none, draw=black,semithick}), for the single-array controller (\addlegendimageintext{mark=none, draw=legreddraw,dashdotted,semithick}) and for the two-array controllers with $\lambda_{\mathrm{s}\backslash\mathrm{f}}=2\pi\times\SI{50}{\radian\per\second}$ (\addlegendimageintext{mark=none, draw=legbluedraw, dashed,semithick}) and $\lambda_{\mathrm{s}\backslash\mathrm{f}}=2\pi\times\SI{10}{\radian\per\second}$ (\addlegendimageintext{mark=none, draw=leggreendraw, dotted,semithick}). The first to fourth rows correspond to BPMs 1, 3, 5 and 7.

To interpret the performance of the single-array controller, consider the Bode magnitude diagram from Fig.~\ref{fig:bodeDIsensitivity}, which shows the output sensitivity of the TISO and SISO systems of the two-array controller. Because the TISO systems are tuned to match the performance of the single-array controller, the TISO output sensitivity from Fig.~\ref{fig:bodeDIsensitivity} (\addlegendimageintext{mark=*, draw=legreddraw,fill=legredfill, dashdotted,semithick}) corresponds to the expected single-array sensitivity \emph{before} including $\Gamma$. From to Fig.~\ref{fig:bodeDIsensitivity}, a maximum attenuation of $\SI{20}{\dB}=0.1$ and $\SI{40}{\dB}=0.01$ is expected at $\SI{10}{\Hz}$ and $\SI{1}{\Hz}$ for disturbances that are aligned to low-order modes of $R$. In Fig.~\ref{fig:psdy_cell1_XY}, these attenuation levels are reflected in BPM 5, but it is worse for other BPMs, in particular those for which the ASD is small for disabled feedback.

As expected from the controller design, the two-array controllers perform worse than the single-array controller, because fewer correctors cover frequencies up $\lambda_{\mathrm{s}\cap\mathrm{f}}=2\pi\times\SI{176}{\radian\per\second}$. However, the second actuator array allows the closed-loop bandwidth to be increased. For frequencies above $\SI{20}{\Hz}$, the performance of the two-array controllers is comparable to the performance of the single-array controller, which suggests that the disturbances are aligned to directions that correspond to the maximum attenuation in Fig.~\ref{fig:bodeDIcompensator:50} and~\ref{fig:bodeDIcompensator:10}. Indeed, the disturbance peaks between $\SI{20}{\Hz}$ to $\SI{100}{\Hz}$ are eigenfrequencies and harmonics of the girders that support the magnets~\cite{IANDIAMONDIORBIT}, which are proportional to the term $R\,d_u(s)$ in~\eqref{eq:distsplit} and therefore aligned with low-order modes. For frequencies below $\SI{20}{\Hz}$, the two-array controllers perform worse than the single-array controller, but from Fig.~\ref{fig:bodeDIcompensator:50} and~\ref{fig:bodeDIcompensator:10}, remain within the theoretical expectations.

Another performance measure is given by the \emph{integrated beam motion} (IBM) on the right-hand side of Fig.~\ref{fig:psdy_cell1_XY}, which is computed from the ASD~\eqref{eq:ASD} as
\begin{align}\label{eq:ibm}
\text{IBM}_{(i)}(\omega_p) = \sqrt{\frac{f_s}{N} \sum_{k=1}^{p} (D_{(i)}(\omega_k))^2}.
\end{align}
The IBM has the effect of smoothing out the ASD, and compared to the left-hand side of Fig.~\ref{fig:psdy_cell1_XY}, the performance difference between the single-array and the two-array controllers largely disappears.
%
\subsubsection{Inputs}
The main reason for augmenting the single-array system~\eqref{eq:CDsystem} with an additional array of actuators is to split the control effort onto two different kinds of corrector magnets: slow but strong magnets that cover low frequencies where the magnitude of the disturbance spectrum is large, and fast but weaker magnets that cover high frequencies where the magnitude of the disturbance spectrum is smaller. 

Fig.~\ref{fig:psd_input} shows the ASD of the inputs ($\si{\ampere\per\sqrt{\Hz}}$) for the experiments from Fig.~\ref{fig:psdy_cell1_XY}. The first row of Fig.~\ref{fig:psd_input} show the two-array controller with $\lambda_{\mathrm{s}\backslash\mathrm{f}}=2\pi\times\SI{50}{\radian\per\second}$, the second row to the two-array controller with $\lambda_{\mathrm{s}\backslash\mathrm{f}}=2\pi\times\SI{10}{\radian\per\second}$, and the third row shows the single-array controller. For the two-array controllers, the first column of Fig.~\ref{fig:psd_input} corresponds to the slow actuators and the second column to the fast actuators.

As expected from the mid-ranging approach, the ASD of the slow correctors is large at low frequencies and rapidly decreases between \SI{10}{\Hz} and \SI{100}{\Hz}. Comparing Fig.~\ref{fig:psd_input_gsvd_slow_v105_13112022_Y} with the transfer functions from $d(s)$ to $u_\mathrm{s}(s)$ and $u_\mathrm{f}(s)$ from Fig.~\ref{fig:bodeDIcompensatoracc:50}, the input gain decreases from roughly \SI{0.1}{\ampere\per\sqrt{\Hz}} (red) at low frequencies to roughly \SI{0.01}{\ampere\per\sqrt{\Hz}} in the \SI{10}{\Hz} to \SI{100}{\Hz} range, which matches the theoretical prediction. Comparing the first row of Fig.~\ref{fig:psd_input} with the second row of Fig.~\ref{fig:psd_input}, it can be seen how lowering the SISO bandwidth from $\lambda_{\mathrm{s}\backslash\mathrm{f}}=2\pi\times\SI{50}{\radian\per\second}$ to $\lambda_{\mathrm{s}\backslash\mathrm{f}}=2\pi\times\SI{10}{\radian\per\second}$ increases the control effort of the fast actuator array. Comparing the input sensitivities from Fig.~\ref{fig:bodeDIcompensatoracc:50} and~\ref{fig:bodeDIcompensatoracc:10}, it can be seen that at \SI{10}{\Hz}, the gain of the two-array controller with $\lambda_{\mathrm{s}\backslash\mathrm{f}}=2\pi\times\SI{50}{\radian\per\second}$ is $\SI{10}{\dB}\approx 0.3$ lower than the gain of the two-array controller with $\lambda_{\mathrm{s}\backslash\mathrm{f}}=2\pi\times\SI{10}{\radian\per\second}$, which is reflected in Fig.~\ref{fig:psd_input_gsvd_fast_v105_13112022_Y} and~\ref{fig:psd_input_gsvd_fast_v106_13112022_Y}.

The slow array of the two-array controllers can also be compared to the single-array controller (Fig.~\ref{fig:psd_input_imc_v109_13112022_Y}). For the single-array controller, it can be seen that a strong control effort is sustained up to \SI{100}{\Hz}, whereas the control of the slow arrays decreases at lower frequencies. For frequencies above \SI{100}{\Hz}, the difference between the slow arrays of the two-array controllers and the single-array controllers is evident.
\section{Conclusion}
In this paper we have proposed a generalised modal decomposition method for the control of two-array CD systems. Our method is based on the GSVD and simultaneously factors the response matrices of each actuator array. In generalised modal space, the two-array system is decoupled into a set of TISO systems and a set of SISO systems. 

Analogous to the single-array case, our two-array controller is designed in generalised modal space using the IMC structure. For systems with $n_\mathrm{f}<n_\mathrm{s}$, an input compensator is added to the IMC structure to account for the non-normal transformation and remove the performance difference between original and generalised modal space. It was shown that the generalised modal decomposition is closely related to the modal decomposition of a hypothetical system with $R=\begin{bmatrix}R_\mathrm{s} & R_\mathrm{f}\end{bmatrix}$, and therefore allows ill-conditioned systems to be treated with regularisation techniques that proved to be efficient for single-array systems. Analogous to the single-array case, the IMC structure was augmented with an output compensator that damps the control action in direction of the small-magnitude singular values of $R$.

In view of the Diamond-II upgrade that will introduce a two-array system, the proposed algorithm was tested on the existing Diamond storage ring. For the implementation, to mimic the Diamond-II situation, the correctors were artificially partitioned into slow and fast arrays that were controlling a subset of the BPMs, and the controller dynamics were designed using mid-ranging control. The two-array controller was compared against a single-array controller, and the results showed that the single-array and the two-array controllers perform similarly well. For the two-array controller, the slow array covered the low frequencies, while the fast array attenuated higher frequencies as intended for the Diamond-II upgrade.

Even though the real-world results proved the feasibility and applicability of the proposed control algorithm, several research questions remain. It was shown that due to the output compensator, certain disturbance directions are amplified at frequencies at which the control action is transferred from one actuator array to the other. This amplification does not occur without output compensator, and future research could focus on modifying the output compensator to avoid disturbance amplification in this particular frequency range.

For designing the controller dynamics, a mid-ranging approach was used. As desired for the Diamond-II upgrade, the mid-ranging approach yields integrating behaviour for the slow actuator array, while the fast actuator array does not contribute to the steady-state control action. However, the mid-ranging approach requires inverting the actuator dynamics $g_\mathrm{s}(s)$ and $g_\mathrm{f}(s)$, but for Diamond-II the dynamics of the fast actuator array may be such that $g_\mathrm{f}(0)=0$. This means that using a mid-ranging approach would result in undesirable integrating behaviour for the fast actuator array. To avoid this problem, one solution would be to invert only parts of $g_\mathrm{f}(s)$ and quantify the resulting performance loss. Alternatively, one could combine the generalised modal decomposition with a $\mathcal{H}_2$ or $\mathcal{H}_\infty$ controller design~\cite[Ch.\ 9.3]{SKOGESTADMULTI}, which would benefit from the sparsity of the system in generalised modal space.

The performance of the algorithms was compared using amplitude spectral density and integrated beam motion plots. While these figures are sufficient to evaluate the performance of a single algorithm, they only allow a partial comparison of different algorithms as the output is subjected to different disturbances when testing the algorithms in practice. As an alternative to these figures, the algorithms could be compared using the output sensitivity, which, in theory, is independent of the actual disturbance affecting the output during experiments. However, the input-output signals that are obtained from the experiments are closed-loop measurements and therefore noise-correlated, which complicates the use of system identification techniques to estimate the output sensitivity. Future research could focus on introducing a beam position reference signal with the aim of identifying the complementary sensitivity. This reference signal would need to cover the whole frequency range during which the control action is significant, as well as the high-dimensional spatial output space. In addition, reference directions that are aligned with higher-order modes would need to be treated differently for ill-conditioned systems.
\appendix
By inspecting Fig.~\ref{fig:imc} for $\Delta(s)=0$, the transfer function from $-y(s)$ to $u(s)$, i.e.\ the standard feedback controller $C:\C^{n_y}\mapsto\C^{n_\mathrm{s}+n_\mathrm{f}}$, is obtained for $\Upsilon=\trans{\begin{bmatrix}I & I\end{bmatrix}}$ as $C(s)=\invbr{I\!-\!Q(s)P(s)}Q(s)\Gamma\!=\!Q(s)\invbr{I\!-\!P(s)Q(s)}\Gamma$,
where the push-through rule~\cite[Ch.\ 3.2]{SKOGESTADMULTI} was used. From the closed-loop dynamics~\eqref{eq:CLorig}, the term $\invbr{I-Q(s)P(s)}$ is
\begin{align*}
\invbr{I-P(s)Q(s)}=X
\diag\left(I\frac{1}{S_{\mathrm{s}\cap\mathrm{f}}(s)},\,I\frac{1}{S_{\mathrm{s}\backslash\mathrm{f}}(s)}\right)
\inv{X},
\end{align*}
so that using~\eqref{eq:inputs} and $Q_{(\cdot)}(s)=U_{(\cdot)}\tilde{Q}_{(\cdot)}(s)\inv{X}$, the standard feedback controller $C(s)$ is obtained as
\begin{align*}
C(s)&=U_{\mathrm{s},\mathrm{f}}
\begin{bmatrix}\inv{\Sigma_\mathrm{s}}q_{\mathrm{s}}(s) & 0 \\ 0 & I q_{\mathrm{s}}(s)\\ \inv{\Sigma_\mathrm{f}} q_\mathrm{s}(s) & 0\end{bmatrix}
\begin{bmatrix}I\frac{1}{S_{\mathrm{s}\cap\mathrm{f}}(s)} & 0\\ 0 & I\frac{1}{S_{\mathrm{s}\backslash\mathrm{f}}(s)}\end{bmatrix}
\inv{X},\\
&=U_{\mathrm{s},\mathrm{f}}
\begin{bmatrix}\frac{1-S_{\mathrm{s}\backslash\mathrm{f}}(s)}{g_\mathrm{s}(s)S_{\mathrm{s}\cap\mathrm{f}}(s)}\inv{\Sigma_\mathrm{s}} & 0\\[1em] 0 & \frac{1-S_{\mathrm{s}\backslash\mathrm{f}}(s)}{g_\mathrm{s}(s)S_{\mathrm{s}\backslash\mathrm{f}}(s)}I\\[1em] \frac{S_{\mathrm{s}\backslash\mathrm{f}}(s)-S_{\mathrm{s}\cap\mathrm{f}}(s)}{g_\mathrm{f}(s)S_{\mathrm{s}\cap\mathrm{f}}(s)}\inv{\Sigma_\mathrm{f}} & 0\end{bmatrix}
\inv{X}\Gamma,
\end{align*}
where $U_{\mathrm{s},\mathrm{f}}\eqdef\diag(U_{\mathrm{s}},U_{\mathrm{f}})$. The open-loop transfer function $L(s)=P(s)C(s)$ is therefore
\begin{align*}
L(s) &= X\begin{bmatrix}\Sigma_\mathrm{s}g_\mathrm{s}(s) & 0 & \Sigma_\mathrm{f}g_\mathrm{f}(s)\\0 & Ig_\mathrm{s}(s) & 0\end{bmatrix}
\trans{U_{\mathrm{s},\mathrm{f}}}C(s),\\
&=X\diag\left(
I\frac{1-S_{\mathrm{s}\cap\mathrm{f}}(s)}{S_{\mathrm{s}\cap\mathrm{f}}(s)},\,\,
I\frac{1-S_{\mathrm{s}\backslash\mathrm{f}}(s)}{S_{\mathrm{s}\backslash\mathrm{f}}(s)}\right)\inv{X}\Gamma.
\end{align*}

\balance
\bibliographystyle{IEEEtran}
{\small
\bibliography{master_bib_abbrev}}

\end{document}